\documentclass[a4paper,12pt]{article}

\usepackage[english]{babel}
\usepackage{graphicx}
\usepackage{colordvi}
\usepackage{fancybox}
\usepackage{cancel}
\usepackage{amsmath,amsfonts, amssymb}
\usepackage{comment}
\usepackage{longtable}
\usepackage{booktabs}
\usepackage[pdftex]{hyperref}
\usepackage{rotating}
\usepackage[hypcap]{caption} 

\usepackage{perpage} 
\MakePerPage{footnote} 

\usepackage{float}
\floatstyle{plaintop}
\restylefloat{table}

\usepackage{color}
\definecolor{gray}{rgb}{0.5,0.5,0.5}

\newcommand\lsim{\mathrel{\rlap{\lower4pt\hbox{\hskip1pt$\sim$}}
    \raise1pt\hbox{$<$}}}
\newcommand\gsim{\mathrel{\rlap{\lower4pt\hbox{\hskip1pt$\sim$}}
    \raise1pt\hbox{$>$}}}

\newcommand{\beq}{\begin{equation}}
\newcommand{\eeq}{\end{equation}}
\newcommand{\bea}{\begin{eqnarray}}
\newcommand{\eea}{\end{eqnarray}}
\newcommand{\bem}{\begin{pmatrix}}
\newcommand{\eem}{\end{pmatrix}}

\def\yd{Y_{27}}
\def\yt{Y_{\overline{351'}}}
\def\yz{\yd\yt^{-1}\yd}
\def\SU{\mathrm{SU}}
\def\SO{\mathrm{SO}}
\def\RED{\textcolor{red}}
\def\BLUE{\textcolor{blue}}

\def\EE{\mathrm{E}_6}
\def\non{\nonumber}

\begin{document}


\begin{flushright}
OSU-HEP-15-01
\end{flushright}

\vspace*{0.8in}

\bigskip

\begin{center}

{\Large\bf A Minimal Supersymmetric $\mathrm{E}_6$ Unified Theory}

\vspace{1cm}

\centerline{
K.S. Babu$^{a,}$\footnote{babu@okstate.edu},
Borut Bajc$^{b,}$\footnote{borut.bajc@ijs.si} and
Vasja Susi\v c$^{b,}$\footnote{vasja.susic@ijs.si}
}
\vspace{0.5cm}
\centerline{$^{a}${\it\small Department of Physics, Oklahoma State University, Stillwater, OK, 74078, USA }}
\centerline{$^{b}$ {\it\small J.\ Stefan Institute, 1000 Ljubljana, Slovenia}}
\end{center}

\bigskip

\begin{abstract}
We show explicitly that supersymmetric $E_6$ Grand Unified Theory with a Higgs sector consisting of $\{27+\overline{27}+351'+\overline{351'}+78\}$ fields provides a realistic scenario for symmetry breaking and fermion mass generation.  While gauge symmetry breaking can be achieved without the
$78$ field, its presence is critical for a successful doublet-triplet mass splitting. The Yukawa sector of the model consists of only two symmetric matrices describing all of quark, lepton and neutrino masses and mixings. The fermion mass matrices are computed at low energy and a fit to the second and third generation masses and mixings is performed. We find a good numerical fit to the low-energy data.  Thus, this model, having $11$ superpotential parameters, alongside the two symmetric Yukawa matrices, seems to be the best realistic candidate for a minimal renormalizable supersymmetric $E_6$ unified theory.
\end{abstract}

\clearpage

\tableofcontents



\section{Introduction}

There are three types of unified groups (simple groups containing the Standard Model gauge group)
that admit chiral representations for fermions: $\SU(N)$ for $N\geq 3$ \cite{Georgi:1974sy},
$\SO(4N+2)$ for $N\ge 2$ and $\EE$ \cite{Gursey:1975ki}. The first two have been thoroughly studied
in many versions, renormalizable as well as non-renormalizable, with and without supersymmetry. Strange enough, $\EE$, the only exceptional group
which contains chiral representations, has been mainly ignored over the last forty years. Apart from a few exceptions where symmetry
breaking was analyzed (via renormalizable potential with $78$ and two $27$s in Ref. \cite{Kalashnikov:1979dq} and via
renormalizable superpotential with $78$, $27$ and $\overline{27}$ in Ref. \cite{Buccella:1987kc}), only the Yukawa sectors
have been touched upon without using explicit constraints arising from symmetry breaking \cite{Gursey:1978fu,Achiman:1978vg,Shafi:1978gg,Ruegg:1979fr,Barbieri:1980vc,Shaw:1980ku}.

Recently, two of us have tried to fill this gap by suggesting a possible minimal renormalizable supersymmetric $\EE$
model~\cite{Bajc:2013qra}. The Higgs sector is composed of pairs of fundamental $27+\overline{27}$ and the two-index symmetric $351'+\overline{351'}$ representations. Much to our surprise we found that, although successful in the symmetry breaking pattern to the Standard Model (SM) gauge group, this sector is unable to provide two light Higgs doublets
of the minimal supersymmetric Standard Model (MSSM), or technically said, to perform the doublet-triplet (DT) splitting.
The impossibility of DT splitting was due to the identical nature of the expressions for a massless doublet and for a
massless triplet; this result was obtained by an explicit computation, and we are not aware of a shortcut explanation of this unusual feature.\footnote{The sparsity of SU(5) breaking vacuum expectation values in the respective mass matrices is likely to be the fundamental origin of this problem.} To put this result in perspective, consider the minimal renormalizable supersymmetric breaking sectors (able to break to the SM group) for other groups. In the minimal renormalizable $\SU(5)$ superpotential (consisting of the adjoint $24$), there are no SM doublets. In the corresponding minimal $\SO(10)$ case (made out of $210$, $126$ and
$\overline{126}$) one has in principle enough doublets, but only one (in $\overline{126}$) is coupled to fermions. In a minimal $\EE$ case
with $27+\overline{27}+351'+\overline{351'}$, there are many more doublets which couple to fermions, yet the naive DT splitting fails (a completely different approach to DT splitting in $\mathrm{E}_6$ can be found in \cite{Maekawa:2002bk}).
We thus proposed in~\cite{Bajc:2013qra} to add another $27+\overline{27}$ pair. In this way, the Yukawa sector consisted of three Yukawa matrices: the matter fields (3 copies of $27_F$) have one coupling to the $\overline{351'}$ and a coupling to each of the two Higgs-like $27$s. Though we did not perform it explicitly, we strongly believe that in this model a realistic fit of the Yukawa sector is possible to obtain, but the
model is essentially not predictive due to the large number of free parameters in the three Yukawa matrices.

The purpose of this paper is to find an $\EE$ theory with only two Yukawa matrices and thus a simpler Yukawa sector than the model in~\cite{Bajc:2013qra}, while keeping the symmetry breaking sector as simple as possible. The idea is to add to the minimal breaking sector (instead of the extra $27+\overline{27}$ pair) another multiplet, which preferably has the following properties:
\begin{enumerate}
\item It does not couple to the matter bilinears (and so is not a $27$, $\overline{351}$ or $\overline{351'}$).
\item It contributes to the symmetry breaking vacuum.
\item It increases the size of the mass matrices for weak doublets and/or color triplets.
\end{enumerate}

The minimal such multiplet is the adjoint representation $78$, for which all of the above properties hold true. We will see that its inclusion is enough to allow the doublet-triplet splitting in the theory,
thus leading to a theory with the correct (SM) vacuum and the correct (MSSM) low energy field theory.

The Yukawa sector in this model consists of two matrices only. Assuming 3 generations, the down quark and charged lepton sectors
have 3 extra vector-like fields, while the neutrino sector has 3 extra vector-like lepton doublet-antidoublet pairs, as well
as 6 SM singlets, all coming from the extra fields in the decomposition of the fundamental 27 into the SM subgroup. The
projection of all these fields into the usual 3 light generations will be performed explicitly and shown, in the simplified case
of only two generations, to provide a realistic fit of the masses and mixing angles.

Special attention needs to be paid to the overall neutrino mass scale. In minimal $\SO(10)$ \cite{Clark:1982ai,Aulakh:1982sw,Aulakh:2003kg}
this scale poses a serious problem and eventually rules out the low-energy supersymmetry scenario
\cite{Aulakh:2005bd,Bajc:2005qe,Aulakh:2005mw,Bertolini:2006pe}: a too low $\SU(2)_R$ breaking is disallowed by
unification constraints and/or $D=6$ proton decay \cite{Bajc:2008dc}, while a too large  breaking scale makes
the right-handed neutrinos too heavy and thus
the seesaw mechanism ineffective to produce a large enough scale for the light neutrinos. In our model this conclusion is avoidable
because there are many possible fields that can break $\SU(2)_R$. While the neutrino mass scale can be adjusted by choosing the vacuum expectation value (VEV) of the $(1,3,10)$ of the Pati-Salam (PS) $\SU(2)_L\times\SU(2)_R\times\SU(4)_C$, located in the $\overline{126}$ of $\SO(10)$, which in turn is found in the $\overline{351'}$ of $\EE$, all other fields' masses depend also on several other VEVs. We can thus arrange a lighter VEV of this $(1,3,10)$
still avoiding dramatic changes in the renormalization group (RG) gauge running.

We arrange the paper in the following way: we define the model and specify the terms in the superpotential in section~\ref{section:define-model}. We find a suitable vacuum solution, which breaks $\EE$ into the SM group in section~\ref{section:SSB}, perform the doublet-triplet splitting in section~\ref{section:DT-splitting} and compute the mass matrices in the Yukawa sector and identify the low energy content of the theory in section~\ref{section:Yukawa-sector}. We then use the obtained results to perform a $2$ generation fit of the masses and mixing angles in section~\ref{section:numeric-fit}. We finish with a discussion in section~\ref{section:discussion}. Five different appendices collect various definitions and
technical results. All explicit computation in $\EE$ has been performed using methods from \cite{Kephart:1981gf} and \cite{Bajc:2013qra}; also, \cite{Anderson:1999em,Anderson:2000ni,Anderson:2001sd} might also prove useful to the reader.

For ease of use we stick to the following color convention in this paper: \RED{red} denotes VEVs coming from the spontaneous symmetry breaking of the unified theory at the scale $M_{GUT}$, while \BLUE{blue} denotes VEVs coming from electroweak (EW) symmetry breaking at the scale $m_{EW}$.

\section{Defining the model\label{section:define-model}}

The renormalizable $E_6$ SUSY GUT  that we consider here is an alternative to the model in~\cite{Bajc:2013qra}. It is motivated by the fact that the minimal Higgs sector in a renormalizable SUSY $\EE$ model, which can break $\EE$ to the Standard Model, is $351'+\overline{351'}+27+\overline{27}$. This minimal breaking sector cannot accommodate doublet-triplet splitting, however, and therefore needs to be extended to get a realistic model. One possible extension is an addition of a $27+\overline{27}$ pair, which was analyzed in~\cite{Bajc:2013qra}, while an alternative, taken in this paper, is to extend it by the representation $78$ instead. The Yukawa sector in the present model will consist of only
two matrices, unlike that in \cite{Bajc:2013qra}, which has three matrices.

Our renormalizable $\EE$ model thus contains the following:
\begin{itemize}
\item The ``fermionic sector'' of three copies of a chiral supermultiplet $27$, denoted by $27_F^i$, with $i=1,2,3$. Also, we assume a $\mathbb{Z}_2$ \textit{matter parity}, under which the $27_F$ are $-1$, and the remaining chiral superfields are $+1$. With this symmetry, the ansatz $\langle 27_F\rangle=0$ is consistent with the equations of motion, which we shall adopt.
\item The ``breaking sector'' consisting of $351'+\overline{351'}+27+\overline{27}+78$.
\end{itemize}

The model under consideration is supersymmetric. The problem of SUSY breaking will not be considered, since it is (usually) an orthogonal problem to breaking the gauge group. One can imagine however, that whatever the detailed mechanism of SUSY breaking might be, we get soft SUSY breaking terms at energies not far above $m_{\textrm{EW}}$, such that we get automatic unification of gauge couplings at $M_{\textrm{GUT}}$ assuming no extra particle states up to the unification scale. The effective low energy theory of our model will thus be MSSM. The soft SUSY breaking terms do not give contributions to the fermion masses at tree level.  A fit to fermion masses and mixings can then be done without specifying the details of
SUSY breaking, although the RG evolution of these parameters does depend on the SUSY scale (which we assume is of order few TeV).

The full superpotential of our model is
\begin{align}
W_{\textrm{full}}&=m_{27}\; 27\;\overline{27}\;+\;m_{351'}\;351'\;\overline{351'}\nonumber \;+\; m_{78}\;78\;78\\
&\quad+\; \lambda_1\; 351'^3 \;+\; \lambda_2\;\overline{351'}^3 \;+\; \lambda_3\;27^2\;\overline{351'} \;+\; \lambda_4\; \overline{27}^2\;351'\nonumber\\
&\quad+\; \lambda_5\;27^3 \;+\; \lambda_6\;\overline{27}^3 \;+\; \lambda_7\;\overline{27}\;78\;27 \;+\; \lambda_8\;\overline{351'}{}\;78\;351'\nonumber \\
&\quad+\; \tfrac{1}{2}\yd^{ij}\;27_F^i\;27^j_F\;27 \;+\; \tfrac{1}{2}\yt^{ij}\;27_F^i\;27^j_F\;\overline{351'}.\label{equation:W-full}
\end{align}

\noindent
Note that the definitions of $351'$ and $\overline{351'}$  used here are switched compared to Slansky's defintions~\cite{Slansky:1981yr}.

\section{Spontaneous symmetry breaking\label{section:SSB}}
\subsection{Equations of motion}

The Higgs sector $351'+\overline{351'}+27+\overline{27}+78$ forms a realistic Higgs sector, which is able to break the gauge group from $E_6$ to $\SU(3)_C\times\SU(2)_L\times\mathrm{U}(1)_Y$. We provide just such a vacuum below.

First, note the decompositions of $27$, $78$ and $351'$ under $\SO(10)\times \mathrm{U}(1)$:
\begin{align}
27&=16(1)+10(-2)+1(4),\\
78&=45(0)+16(-3)+\overline{16}(3)+1(0),\\
351'&=1(8)+10(2)+16(5)+54(-4)+126(2)+\overline{144}(-1).
\end{align}

The representation $351'$ contains $5$ SM singlets, $3$ of which are $\SU(5)$ singlets (in $1$, $16$ and $126$ of $\SO(10)$), and $2$ are part of a $24$ under $\SU(5)$ (in $54$ and $\overline{144}$ of $\SO(10)$). Similarly, the representation $78$ also has $5$ singlets, with one being a $24$ under $\SU(5)$ (the one in $45$ of $\SO(10)$), while the remaining ones are singlets under $\SU(5)$ (the $1$, $16$ and $\overline{16}$ of $\SO(10)$, as well as another one in the $45$). The Higgs sector $351'+\overline{351'}+27+\overline{27}+78$ therefore contains $5+5+2+2+5=19$ singlets in total. We list their VEVs\footnote{
Notice that the Standard Model singlet VEVs have been denoted by $u_1$, $u_2$, $v$, $w$ and $y$. The notation from \cite{Buccella:1987kc} is changed due to the states now being those which have well defined transformation properties under the $\SU(5)$ and $\SO(10)$ subgroups of $\EE$. The connection between the two notations is
$u_1\propto a_1$, $u_2\propto a_2$, $w\propto a_3-b_3$, $v\propto -\sqrt{3}a_3+2a_4-\sqrt{3}b_3$,
$y\propto a_3+\sqrt{3}a_4+b_3$, with the usual normalization
$\langle 78^{i}{}_{j} 78^{\ast}_ i{}^{j}\rangle=|u_1|^2+|u_2|^2+|v|^2+|w|^2+|y|^2$.} in Table~\ref{table:singlet-labels}.

\begin{table}[h]
\caption{SM singlet VEVs in our Higgs sector.\label{table:singlet-labels}}
\vskip 0.2cm
\centering
\scalebox{0.9}{
\begin{tabular}{rrrrr@{\hspace{1.5cm}}rrrrr}
\toprule
label&
$\subseteq\mathrm{PS}$&
$\subseteq\SU(5)$&
$\subseteq\SO(10)$&
$\subseteq \mathrm{E}_6$&
label&
$\subseteq\mathrm{PS}$&
$\subseteq\SU(5)$&
$\subseteq\SO(10)$&
$\subseteq \mathrm{E}_6$\\\midrule
$c_1$&$(1,1,1)$&      $1$    &$1$                &$27$&$d_1$&$(1,1,1)$&$1$    &$1$                &$\overline{27}$\\
$c_2$&$(1,2,\overline{4})$&      $1$    &$16$               &$27$&$d_2$&$(1,2,4)$&$1$    &$\overline{16}$    &$\overline{27}$\\\addlinespace
$e_1$&$(1,3,10)$&$1$&$126$&$351'$&$f_1$&$(1,3,\overline{10})$&$1$&$\overline{126}$&$\overline{351'}$\\
$e_2$&$(1,2,\overline{4})$&$1$&$16$&$351'$&$f_2$&$(1,2,4)$&$1$&$\overline{16}$&$\overline{351'}$\\
$e_3$&$(1,1,1)$&$1$&$1$&$351'$&$f_3$&$(1,1,1)$&$1$&$1$&$\overline{351'}$\\
$e_4$&$(1,1,1)$&$24$&$54$&$351'$&$f_4$&$(1,1,1)$&$24$&$54$&$\overline{351'}$\\
$e_5$&$(1,2,4)$&$24$&$\overline{144}$&$351'$&$f_5$&$(1,2,\overline{4})$&$24$&$144$&$\overline{351'}$\\\addlinespace
$u_1$&$(1,2,\overline{4})$&$1$&$16$&$78$&&&&&\\
$u_2$&$(1,2,4)$&$1$&$\overline{16}$&$78$&&&&&\\
$w_{\phantom{1}}$&$(1,1,1)$&$1$&$1$&$78$&&&&&\\
$v_{\phantom{1}}$&$/$&$1$&$45$&$78$&&&&&\\
$y_{\phantom{1}}$&$/$&$24$&$45$&$78$&&&&&\\\bottomrule
\end{tabular}
}
\end{table}

With the ansatz $\langle 27_F\rangle=0$, the breaking part of the superpotential is
\begin{align}
W&=m_{351'}\;I_{351'\times\overline{351'}}+m_{27}\;I_{27\times\overline{27}}+m_{78}\;I_{78\times 78}\nonumber\\
&\qquad + \lambda_1 \;I_{351'^3}+\lambda_2 \;I_{\overline{351'}^3}+\lambda_3 \;I_{27^2\times \overline{351'}}+\lambda_4 \;I_{\overline{27}^2\times 351'}\nonumber\\
&\qquad +\lambda_5 \;I_{27^3} + \lambda_6 \;I_{\overline{27}^3}+\lambda_7\; I_{27\times 78\times \overline{27}}+\lambda_8\; I_{351'\times 78\times \overline{351'}}\label{equation:superpotential-breaking-sector}.
\end{align}

Explicit computation yields the following expressions for the superpotential invariants (VEV terms only):
        \begin{align}
        I_{351'\times\overline{351'}}&=\overline{351'}_{\mu\nu}\;351^{\mu\nu}=\RED{e_1} \RED{f_1}+\RED{e_2} \RED{f_2}+\RED{e_3} \RED{f_3}+\RED{e_4} \RED{f_4}+\RED{e_5} \RED{f_5},\label{equation:computed-invariant-first}\\
        I_{27\times\overline{27}}&=\overline{27}_\mu\; 27^\mu=\RED{c_1} \RED{d_1}+\RED{c_2} \RED{d_2},\\
        I_{78\times 78}&=78^{\mu}{}_{\nu}\;78^{\nu}{}_{\mu}=2 \RED{u_1} \RED{u_2} + \RED{w}^2+\RED{v}^2+\RED{y}^2,\\
        I_{351'^3}&=351'^{\mu\alpha}\;351'^{\nu\beta}\;351'^{\lambda\gamma}\;d_{\alpha\beta\gamma}d_{\mu\nu\lambda}=3 \left(\RED{e_3} \RED{e_4}^2+\RED{e_1} \RED{e_5}^2 - \sqrt{2} \RED{e_2} \RED{e_4} \RED{e_5} \right),\\
        I_{\overline{351'}^3}&=\overline{351'}_{\mu\alpha}\;\overline{351'}_{\nu\beta}\;\overline{351'}_{\lambda\gamma}\;d^{\alpha\beta\gamma}\;d^{\mu\nu\lambda}=3 \left(\RED{f_3} \RED{f_4}^2+\RED{f_1} \RED{f_5}^2 - \sqrt{2}\RED{f_2} \RED{f_4} \RED{f_5}\right),\\
        I_{27^2\times\overline{351'}}&=\overline{351'}_{\mu\nu}\;27^\mu\;27^\nu=\RED{c_2}^2 \RED{f_1} + \sqrt{2} \RED{c_1} \RED{c_2} \RED{f_2} + \RED{c_1}^2 \RED{f_3},\\
        I_{\overline{27}^2\times 351'}&=351'^{\mu\nu}\;\overline{27}_\mu\;\overline{27}_\nu=\RED{d_2}^2 \RED{e_1} + \sqrt{2} \RED{d_1} \RED{d_2} \RED{e_2} + \RED{d_1}^2 \RED{e_3},\\
        I_{27^3}&=27^\mu\;27^\nu\;27^\lambda\;d_{\mu\nu\lambda}=0,\\
        I_{\overline{27}^3}&=\overline{27}_\mu\;\overline{27}_\nu\;\overline{27}_\lambda\;d^{\mu\nu\lambda}=0,\\
        I_{27\times 78\times \overline{27}}&=\overline{27}_{\mu}\;78^{\mu}{}_{\nu}\;27^\nu\nonumber\\
        &=
        \tfrac{1}{\sqrt{6}}\RED{u_1}\RED{c_1}\RED{d_2} + \tfrac{1}{\sqrt{6}}\RED{u_2} \RED{c_2} \RED{d_1}-\tfrac{1}{6\sqrt{2}}\RED{w}\left(4 \RED{c_1}\RED{d_1}+\RED{c_2}\RED{d_2}\right)-\tfrac{1}{2} \sqrt{\tfrac{5}{6}} \RED{v} \RED{c_2}\RED{d_2},\\
        I_{351'\times 78\times \overline{351'}}&=\overline{351'}_{\mu\nu}\;78^{\nu}{}_{\kappa}\;351'^{\kappa\mu}=\nonumber\\
        &=\tfrac{1}{2\sqrt{6}}\RED{u_1}\left(\sqrt{2} \RED{e_2} \RED{f_1}+\sqrt{2} \RED{e_3} \RED{f_2}+\RED{e_5} \RED{f_4}\right)+
        \tfrac{1}{2\sqrt{6}}\RED{u_2}\left(\sqrt{2} \RED{e_1} \RED{f_2}+\sqrt{2} \RED{e_2} \RED{f_3}+\RED{e_4} \RED{f_5}\right)\nonumber\\
        &\quad +
        \tfrac{1}{12\sqrt{2}}\RED{w}\left(-2 \RED{e_1} \RED{f_1}-5 \RED{e_2} \RED{f_2}-8 \RED{e_3} \RED{f_3}+4 \RED{e_4} \RED{f_4}+\RED{e_5} \RED{f_5}\right)
        -\tfrac{1}{4} \sqrt{\tfrac{5}{6}} \RED{v}\left(2 \RED{e_1} \RED{f_1}+\RED{e_2} \RED{f_2}-\RED{e_5} \RED{f_5}\right).\label{equation:computed-invariant-last}
        \end{align}

The general formula for $D$-terms in our case yields

\begin{align}
D^A&= \phantom{+}(27^\dagger)_\mu\;(\hat{t}^A\, 27)^\mu+(\overline{27}^\dagger)^\mu\;(\hat{t}^A\, \overline{27})_\mu+(78^\dagger)^{\nu}{}_{\mu}\;(\hat{t}^A\, 78)^{\mu}{}_{\nu}\nonumber\\
&\quad+(351'^\dagger)_{\mu\nu}\;(\hat{t}^A\, 351')^{\mu\nu}+(\overline{351'}^\dagger)^{\mu\nu}\;(\hat{t}^A\, \overline{351'})_{\mu\nu}.
\end{align}

Of the $78$ $D$-terms, $5$ vanish non-trivially, corresponding to the following generators of the $\SU(3)_C\times\SU(3)_L\times\SU(3)_R$ subgroup of $\EE$: $t_L^8$, $t_R^3$, $t_R^6$, $t_R^7$, $t_R^8$, given explicitly by

\begin{align}
D_L^8&=\tfrac{1}{\sqrt{3}}\left(|\RED{c_1}|^2+|\RED{c_2}|^2+2 |\RED{e_1}|^2 +2 |\RED{e_2}|^2 +2 |\RED{e_3}|^2 -|\RED{e_4}|^2 -|\RED{e_5}|^2\right.\nonumber\\
&\qquad \left.-|\RED{d_1}|^2 -|\RED{d_2}|^2-2 |\RED{f_1}|^2 -2 |\RED{f_2}|^2 -2 |\RED{f_3}|^2 +|\RED{f_4}|^2 +|\RED{f_5}|^2\right),\\
D_R^3&=\tfrac{1}{6}\left(
-3 |\RED{c_2}|^2 -6 |\RED{e_1}|^2 -3 |\RED{e_2}|^2 +3 |\RED{e_5}|^2 -|\RED{u_1}|^2 \right.\nonumber\\
&\qquad\left.+3 |\RED{d_2}|^2 +6 |\RED{f_1}|^2 +3 |\RED{f_2}|^2 -3 |\RED{f_5}|^2 +|\RED{u_2}|^2\right),\\
D_R^6&=
\tfrac{1}{12} \left(
6 \RED{c_2} \RED{c_1}^\ast+6 \RED{c_1} \RED{c_2}^\ast
-\sqrt{3} \RED{u_1} \RED{w}^\ast-\sqrt{3}\RED{w} \RED{u_1}^\ast
+\sqrt{5} \RED{u_1} \RED{v}^\ast +\sqrt{5} \RED{v} \RED{u_1}^\ast\right.\nonumber\\
&\qquad\left.-6 \RED{d_2} \RED{d_1}^\ast-6 \RED{d_1} \RED{d_2}^\ast
+\sqrt{3} \RED{u_2} \RED{w}^\ast+\sqrt{3}\RED{w} \RED{u_2}^\ast
-\sqrt{5}\RED{u_2} \RED{v}^\ast -\sqrt{5} \RED{v} \RED{u_2}^\ast\right.\nonumber\\
&\qquad\left.+6\sqrt{2} \RED{e_1} \RED{e_2}^\ast+6\sqrt{2}\RED{e_2} \RED{e_1}^\ast
+6\sqrt{2} \RED{e_2} \RED{e_3}^\ast+6\sqrt{2}\RED{e_3} \RED{e_2}^\ast
+6 \RED{e_4} \RED{e_5}^\ast+6 \RED{e_5} \RED{e_4}^\ast\right.\nonumber\\
&\qquad \left. -6\sqrt{2} \RED{f_1}\RED{f_2}^\ast-6\sqrt{2}\RED{f_2}\RED{f_1}^\ast
-6\sqrt{2} \RED{f_2}\RED{f_3}^\ast-6\sqrt{2}\RED{f_3}\RED{f_2}^\ast
-6 \RED{f_4} \RED{f_5}^\ast-6 \RED{f_5} \RED{f_4}^\ast \right),\\
D_R^7&=
\tfrac{i}{12} \left(
6 \RED{c_2} \RED{c_1}^\ast-6 \RED{c_1} \RED{c_2}^\ast
-\sqrt{3} \RED{u_1} \RED{w}^\ast+\sqrt{3}\RED{w} \RED{u_1}^\ast
+\sqrt{5} \RED{u_1} \RED{v}^\ast -\sqrt{5} \RED{v} \RED{u_1}^\ast\right.\nonumber\\
&\qquad\left.+6 \RED{d_2} \RED{d_1}^\ast-6 \RED{d_1} \RED{d_2}^\ast
-\sqrt{3} \RED{u_2} \RED{w}^\ast+\sqrt{3}\RED{w} \RED{u_2}^\ast
+\sqrt{5}\RED{u_2} \RED{v}^\ast -\sqrt{5} \RED{v} \RED{u_2}^\ast\right.\nonumber\\
&\qquad\left.+6\sqrt{2} \RED{e_1} \RED{e_2}^\ast-6\sqrt{2}\RED{e_2} \RED{e_1}^\ast
+6\sqrt{2} \RED{e_2} \RED{e_3}^\ast-6\sqrt{2}\RED{e_3} \RED{e_2}^\ast
+6 \RED{e_4} \RED{e_5}^\ast-6 \RED{e_5} \RED{e_4}^\ast\right.\nonumber\\
&\qquad \left. +6\sqrt{2} \RED{f_1}\RED{f_2}^\ast-6\sqrt{2}\RED{f_2}\RED{f_1}^\ast
+6\sqrt{2} \RED{f_2}\RED{f_3}^\ast-6\sqrt{2}\RED{f_3}\RED{f_2}^\ast
+6 \RED{f_4} \RED{f_5}^\ast-6 \RED{f_5} \RED{f_4}^\ast \right),\\
D_R^8&=\tfrac{1}{2\sqrt{3}}\left(-2 |\RED{c_1}|^2 +|\RED{c_2}|^2 +2 |\RED{e_1}|^2 -|\RED{e_2}|^2 -4 |\RED{e_3}|^2 +2 |\RED{e_4}|^2 -|\RED{e_5}|^2 +|\RED{u_1}|^2\right.\nonumber\\
&\qquad\quad\left. +2 |\RED{d_1}|^2 -|\RED{d_2}|^2 -2 |\RED{f_1}|^2 +|\RED{f_2}|^2 +4 |\RED{f_3}|^2 -2 |\RED{f_4}|^2 +|\RED{f_5}|^2 -|\RED{u_2}|^2
\right).
\end{align}

They can be rewritten into $3$ independent $D$-terms:

\begin{align}
D^{I}&=|\RED{c_1}|^2 - |\RED{d_1}|^2+ |\RED{e_2}|^2 -|\RED{f_2}|^2 +2 |\RED{e_3}|^2 -2 |\RED{f_3}|^2 -|\RED{e_4}|^2 +|\RED{f_4}|^2 -\tfrac{1}{3}|\RED{u_1}|^2+\tfrac{1}{3}|\RED{u_2}|^2,\label{equation:D-term-summary-first}\\
D^{II}&=|\RED{c_2}|^2-|\RED{d_2}|^2+|\RED{e_2}|^2-|\RED{f_2}|^2+2 |\RED{e_1}|^2-2 |\RED{f_1}|^2-|\RED{e_5}|^2+|\RED{f_5}|^2-\tfrac{1}{3}|\RED{u_2}|^2+\tfrac{1}{3}|\RED{u_1}|^2,\\
D^{III}&=+
\RED{c_1} \RED{c_2}^\ast-\tfrac{\sqrt{3}}{6} \RED{w} \RED{u_1}^\ast+\tfrac{\sqrt{5}}{6} \RED{v} \RED{u_1}^\ast
+\sqrt{2} \RED{e_2}\RED{e_1}^\ast+\sqrt{2} \RED{e_3}\RED{e_2}^\ast+\RED{e_5}\RED{e_4}^\ast\nonumber\\
&\quad - \RED{d_2}\RED{d_1}^\ast+\tfrac{\sqrt{3}}{6} \RED{u_2} \RED{w}^\ast-\tfrac{\sqrt{5}}{6} \RED{u_2} \RED{v}^\ast
-\sqrt{2} \RED{f_1}\RED{f_2}^\ast-\sqrt{2}\RED{f_2}\RED{f_3}^\ast-\RED{f_4}\RED{f_5}^\ast,\label{equation:D-term-summary-last}
\end{align}

\noindent
via the definitions $D^I:=\sqrt{3}D_L^8+2D_R^3$, $D^{II}:=-2D_R^3$ and $D^{III}:=D_R^6+i D_R^7$, where $D^{III}$ now forms a complex equation. The other independent combination $D^Y=D_L^8+\sqrt{3}D_R^3+D_R^8$ is trivially zero, as it should be, since this $D$-term corresponds to the unbroken generator of $\mathrm{U}(1)_Y$.

\subsection{A specific vacuum solution}
In this section, we obtain a vacuum solution which breaks the gauge group to the Standard Model group $\SU(3)_C\times\SU(2)_L\times\mathrm{U}(1)_Y$. Due to the complexity of the equations we are unable to provide a full classification of vacua; a short discussion on alternative vacua can be found in Appendix~\ref{section:vacuum2}.

The equations of motion in SUSY models are
\begin{align}
D^a&=0,\\
F_\phi&=0,
\end{align}
with the usual definition of the $F$-term:
\begin{align}
F_\phi:=\frac{\partial W}{\partial \phi}.
\end{align}

In our case, there are $19$ Standard Model singlets in the Higgs sector, giving $19$ non-trivial $F$-terms, which can be easily reconstructed from the superpotential in equation~\eqref{equation:superpotential-breaking-sector} and the all-singlet terms of the invariants given by equations~\eqref{equation:computed-invariant-first}--\eqref{equation:computed-invariant-last}. The non-trivial $D$-terms are given by equations~\eqref{equation:D-term-summary-first}--\eqref{equation:D-term-summary-last}.

To obtain a vacuum solution, we perform the following steps:
\begin{itemize}
\item
First, we notice that $F_y$ leads directly to $y=0$ ($y$ is present only in the mass term $78^2$). By taking the self-consistent ansatz\footnote{This ansatz is motivated by the ansatz used in the model with the $78$ omitted~\cite{Bajc:2013qra}, where a classification of vacua is known. See Appendix~\ref{section:vacuum2} for further discussion.}
    \begin{align}
    c_1=d_1=f_5=e_5&=0,\label{equation:ansatz-first}\\
    u_1=u_2=e_2=f_2&=0,\label{equation:ansatz-last}
    \end{align}

the system of equations is greatly simplified: $F_{c_1},F_{d_1},F_{e_2},F_{f_2},F_{e_5},F_{f_5},F_{u_1},F_{u_2},F_y$ and $D^{III}$ are solved automatically.
\item Solve $F_{e_3}$ and $F_{f_3}$ for $f_3$ and $e_3$, respectively to get
    \begin{align}
    e_3&=-\frac{9 \lambda_2 \RED{f_4}^2}{3 m_{351'}-\sqrt{2} \lambda_8 \RED{w}},&
    f_3&=-\frac{9 \lambda_1 \RED{e_4}^2}{3 m_{351'}-\sqrt{2} \lambda_8 \RED{w}}.
    \end{align}
\item Solve $F_{c_2}$ and $F_{d_2}$ for $f_1$ and $e_1$, respectively to get
    \begin{align}
    e_1&=\frac{\RED{c_2}}{24 \lambda_4 \RED{d_2}}\left(\sqrt{2} \lambda_7 \left(\sqrt{15} \RED{v}+\RED{w}\right)-12 m_{27} \right),\\
    f_1&=\frac{\RED{d_2}}{24 \lambda_3 \RED{c_2}}\left(\sqrt{2} \lambda_7 \left(\sqrt{15} \RED{v}+\RED{w}\right)-12 m_{27} \right).
    \end{align}
\item Simultaneously solve $F_{e_4}$ and $F_{f_4}$ for $f_4$:
    \begin{align}
    f_4&=\frac{(3 \sqrt{2} m_{351'}-2 \lambda_8 \RED{w})(3 \sqrt{2} m_{351'}+\lambda_8\RED{w})}{324 \RED{e_4} \lambda _1 \lambda _2}.
    \end{align}
\item Simultaneously solve $F_{e_1}$ and $F_{f_1}$ for $d_2$:
    \begin{align}
    d_2&=\frac{\big(6\sqrt{2}m_{27}-\lambda_7(\sqrt{15}\RED{v}+\RED{w})\big)\big(6\sqrt{2}m_{351'}-\lambda_8 (\sqrt{15}\RED{v}+\RED{w})\big)}{144 \lambda_3 \lambda_4 \RED{c_2}}.
    \end{align}
\item It is now convenient to define a new quantity $A:=\sqrt{15} v+ w$. We can now solve $F_{v}$ for $v$ as a linear equation:
    \begin{align}
    v&=\frac{\sqrt{10} \RED{A}^2 \lambda_7^2 \lambda_8-8 \sqrt{5} \lambda_7\RED{A} (m_{351'}\lambda_7 +2 m_{27}\lambda_8)+24 \sqrt{10} m_{27} (2 m_{351'} \lambda_7+m_{27} \lambda_8)}{768 \sqrt{3} m_{78} \lambda_3 \lambda_4}.
    \end{align}
\item Three variables remain to be determined: $A$, $c_2$ and $e_4$. We are left with only one unsolved $F$-term $F_{w}$, which is a polynomial in $A$:
    \begin{align}
    0&=P_0+P_1 \RED{A}+P_2 \RED{A}^2+P_3 \RED{A}^3+P_4 \RED{A}^4,\\
    P_0&=-576 m_{27} \left(2 m_{351'} \lambda_7+m_{27} \lambda_8\right)\Big(25 m_{27} \left(2 m_{351'} \lambda_7+m_{27} \lambda_8\right) \lambda_8^3\nonumber\\
    &\qquad -480 m_{351'} m_{78} \lambda_3
   \lambda_4 \lambda_8^2+110592 m_{78}^2 \lambda_1 \lambda_2 \lambda_3 \lambda_4\Big),\\
   P_1&=192\sqrt{2}\Big(995328 \lambda_1 \lambda_2 \lambda_3^2 \lambda_4^2 m_{78}^3\nonumber\\
   &\qquad +4608 \lambda_3 \lambda_4 \left(24 m_{27} \lambda_1 \lambda_2 \lambda_7 \lambda_8+m_{351'} \left(12 \lambda
   _1 \lambda_2 \lambda_7^2-\lambda_3 \lambda_4 \lambda_8^2\right)\right) m_{78}^2\nonumber\\
   &\qquad +240 \lambda_3 \lambda_4 \lambda_8^2 \left(-m_{351'}^2 \lambda_7^2+2 m_{351'} m_{27} \lambda
   _8 \lambda_7+2 m_{27}^2 \lambda_8^2\right) m_{78}\nonumber\\
   &\qquad+25 m_{27} \lambda_7 \lambda_8^3 \left(2 m_{351'}^2 \lambda_7^2+5 m_{351'} m_{27} \lambda_8 \lambda_7+2 m_{27}^2 \lambda_8^2\right)\Big),\\
   P_2&=-16\Big(\lambda _8 \left(18432 \lambda _3 \lambda _4 \left(9 \lambda _1 \lambda _2 \lambda _7^2+\lambda _3 \lambda _4 \lambda _8^2\right) m_{78}^2\right.\nonumber\\
   &\qquad \left.+240 \lambda _3 \lambda _4 \lambda
   _7 \lambda _8^2 \left(5 m_{351'} \lambda _7+16 m_{27} \lambda _8\right) m_{78}\right.\nonumber\\
   &\qquad \left.+25 \lambda _7^2 \lambda _8^2 \left(2 m_{351'}^2 \lambda _7^2+14 m_{351'} m_{27} \lambda _8 \lambda
   _7+11 m_{27}^2 \lambda _8^2\right)\right)\Big),\\
   P_3&=40\sqrt{2}\lambda _7^2 \lambda _8^4 \left(96 m_{78} \lambda _3 \lambda _4+5 \lambda _7 \left(m_{351'} \lambda _7+2 m_{27} \lambda _8\right)\right),\\
   P_4&=-25\lambda _7^4 \lambda _8^5.
    \end{align}

Note that the coefficients $P_i$ depend only on the Lagrangian parameters; choosing those, we can determine $A$ numerically.

Finally, we solve the remaining $D$-terms and determine $c_2$ and $e_4$ from
\begin{align}
0&=D^{I}=2 |\RED{e_3}|^2-|\RED{e_4}|^2-2 |\RED{f_3}|^2+|\RED{f_4}|^2,\\
0&=D^{II}=|\RED{c_2}|^2-|\RED{d_2}|^2+2 |\RED{e_1}|^2-2 |\RED{f_1}|^2.
\end{align}
Note that $v$ and $w$ are determined, once $A$ is fixed. We can therefore see that $f_4\propto 1/e_4$, $e_3\propto f_4^2\propto 1/e_4^2$ and $f_3\propto e_4^2$. Therefore $D^{I}$ can be written as a quartic polynomial in $|e_4|^2$; the constant term has a positive coefficient (the $e_3$ term), while the highest order term in $|e_4|^2$ has a negative coefficient, a solution for a real $e_4>0$ will always exist. Similarly,
$d_2\propto 1/c_2$, $e_1\propto c_2/d_2\propto c_2^2$ and $f_1\propto d_2/c_2\propto 1/c_2^2$, $D^{II}$ is a quartic polynomial in $c_2$ independent of $e_4$; the constant coefficient will be negative (the $f_1$ term), while the highest order coefficient in $|c_2|^2$ is positive (the $e_1$ term), which again guarantees a real solution $c_2>0$.

\end{itemize}

The initial Lagrangian parameters constitute of the $3$ masses $m_{27}$, $m_{351'}$, $m_{78}$ and $8$ massless parameters $\lambda_i$, $i=1,\ldots,8$. The simplest order in which to compute the given vacuum solution with these parameters is given below:
\begin{enumerate}
\item Take the ansatz for some of the VEVs (symmetric under conjugation symmetry \cite{Bajc:2013qra})
\begin{align}
c_1&=0,&d_1&=0,\non\\
e_2&=0,&f_2&=0, \non\\
e_5&=0,&f_5&=0, \non\\
u_1&=0,&u_2&=0, \non\\
y&=0.&&
\end{align}
\item $A$ is determined through the polynomial, and then the VEVs $v$ and $w$ are determined by
\begin{align}
v&=\frac{\sqrt{10} \RED{A}^2 \lambda_7^2 \lambda_8-8 \sqrt{5} \lambda_7\RED{A} (m_{351'}\lambda_7 +2 m_{27}\lambda_8)+24 \sqrt{10} m_{27} (2 m_{351'} \lambda_7+m_{27} \lambda_8)}{768 \sqrt{3} m_{78} \lambda_3 \lambda_4},\\
w&=\RED{A}+\frac{5}{768 m_{78} \lambda_3 \lambda_4}\left(-24 \sqrt{2} m_{27} (2 m_{351'} \lambda_7+m_{27}
   \lambda_8)\right.\nonumber\\
&\qquad \left.+\lambda_7\RED{A}(8 m_{351'} \lambda_7+16 m_{27}
   \lambda_8-\sqrt{2} \RED{A} \lambda_7 \lambda_8)\right).
\end{align}
\item
$e_4$ and $c_2$ are determined through $D^{I}$ and $D^{II}$, respectively, using equations~\eqref{equation:VEVan-1}--\eqref{equation:VEVan-5}.
\item The remaining nonvanishing VEVs are $d_2,e_1,f_1,e_3,f_3,f_4$, and they can be computed in terms of $A,w,c_2,e_4$:
\begin{align}
d_2&=\frac{1}{144 \lambda_3 \lambda_4 \RED{c_2}}\left(6\sqrt{2}m_{27}-\lambda_7\RED{A}\right)\left(6\sqrt{2}m_{351'}-\lambda_8 \RED{A}\right),\label{equation:VEVan-1}\\
e_1&=\frac{12 \lambda_3 \RED{c_2}^2}{\sqrt{2} \text{\RED{A}} \lambda_8-12 m_{351'}},\\
f_1&=\frac{\sqrt{2} \RED{A}\lambda_7-12 m_{27}}{3456 \lambda_3^2 \lambda_4
   \RED{c_2}^2}\left(6\sqrt{2}m_{27}-\lambda_7\RED{A}\right)\left(6\sqrt{2}m_{351'}-\lambda_8 \RED{A}\right),\\
e_3&=\frac{\left(3 \sqrt{2} m_{351'}-2 \lambda_8 \RED{w}\right) \left(3 \sqrt{2} m_{351'}+\lambda_8\RED{w}\right)^2}{5832 \sqrt{2} \RED{e_4}^2 \lambda_1^2 \lambda_2},\\
f_3&=-\frac{9 \RED{e_4}^2 \lambda _1}{3 m_{351'}-\sqrt{2} \RED{w} \lambda _8},\\
f_4&=\frac{1}{324 \RED{e_4} \lambda _1 \lambda _2}\left(3 \sqrt{2} m_{351'}-2 \lambda_8 \RED{w}\right)\left(3 \sqrt{2} m_{351'}+\lambda_8\RED{w}\right).\label{equation:VEVan-5}
\end{align}
\end{enumerate}

The solution above does indeed break $\EE$ into the Standard Model group. This can be checked by explicitly computing the gauge boson masses, found in Table~\ref{table:gauge-boson-masses} of Appendix~\ref{Gauge-Boson-masses}. It is possible to further illuminate this result by considering that under the standard embeddings of the $\EE$ subgroups from Table~\ref{table:singlet-labels}, the $\SU(5)$ breaking in our solution is solely due to non-vanishing VEVs $e_4$ and $f_4$.

\section{Doublet-triplet splitting\label{section:DT-splitting}}

We tackle now the issue of doublet-triplet splitting. We denote the doublets and antidoublets by $D\sim (1,2,+1/2)$ and $\overline{D}\sim (1,2,-1/2)$, while the triplets and antitriplets are denoted by $T\sim (3,1,-1/3)$ and $\overline{T}\sim (\bar{3},1,+1/3)$. The detailed labels of these states are given in Table~\ref{table:labels-doublets-triplets} of Appendix \ref{Particle-identification}.

In the Higgs sector of our model, there are $12$ doublet-antidoublet pairs and $13$ triplet-antitriplet pairs. This is one extra pair of each compared to the renormalizable model with the Higgs sector $351'+\overline{351'}+27+\overline{27}$, in which doublet-triplet splitting surprisingly fails (see~\cite{Bajc:2013qra} for details). The extra states come from the added $78$, and are labeled by the index $0$, i.e.~$D_0,\overline{D}_0,T_0,\overline{T}_0$. We shall see that  this extra row and column, together with a new vacuum compared to the model without $78$ now enable doublet-triplet splitting in the usual way (by fine-tuning). Note that all the doublets and triplets are located in a $5$ or $45$ (or their conjugates) of $\SU(5)$, except for one extra triplet in the $50$ of $\SU(5)$.

The mass terms for the doublets and triplets are written as

\begin{align}
\begin{pmatrix}D_0&\cdots& D_{11}\\\end{pmatrix}
 \mathcal{M}_{\textrm{doublets}}
\begin{pmatrix}\overline{D}_1\\\vdots \\\overline{D}_{11}\\\end{pmatrix}+
\begin{pmatrix}T_0&\cdots& T_{12}\\\end{pmatrix}
 \mathcal{M}_{\textrm{triplets}}
\begin{pmatrix}\overline{T}_1\\\vdots \\\overline{T}_{12}\\\end{pmatrix}.
\end{align}

The matrices $\mathcal{M}_{\textrm{doublets}}$ and $\mathcal{M}_{\textrm{triplets}}$ are similar; we can compactly write a $13\times 13$ matrix $\mathcal{M}$ with block form
\begin{align}
\mathcal{M}&=
\begin{pmatrix}
M_{11}&M_{12}\\
M_{21}&M_{22}\\
\end{pmatrix},\label{equation:DT-matrix}
\end{align}
with the diagonal blocks defined by
\begin{align}
M_{11}&=\scalebox{0.7}{$
\left(\begin{smallmatrix}
m_{78} & 0 & \frac{\lambda_{7} \RED{d_2}}{2 \sqrt{3}} & 0 & 0 & 0 \\
 0 & m_{27}+\frac{\lambda_{7} (\sqrt{3}\RED{v}+\sqrt{5}\RED{w})}{3\sqrt{10}}& \frac{\alpha  \lambda_{3} \RED{f_4}}{\sqrt{15}} & 6 \lambda_{5} \RED{c_2} & 0 & 0\\
 \frac{\lambda_{7} \RED{c_2}}{2 \sqrt{3}} & \frac{\alpha  \lambda_{4} \RED{e_4}}{\sqrt{15}} & m_{27}+\frac{\lambda_{7} (\sqrt{5}\RED{w}-\sqrt{3}\RED{v})}{3\sqrt{10}} & 0 & 0 & 0 \\
 0 & 6 \lambda_{6} \RED{d_2} & 0 & m_{27}+\frac{\lambda_7 (3\sqrt{3}\RED{v}-\sqrt{5}\RED{w})}{6\sqrt{10}}& 0 & -\frac{\lambda_{4} \RED{d_2}}{\sqrt{10}} \\
 0 & 0 & 0 & 0 & m_{351'}+\frac{\lambda_{8} \RED{v}}{2 \sqrt{30}}-\frac{\lambda_{8} \RED{w}}{6 \sqrt{2}} & \sqrt{\frac{3}{5}} \alpha  \lambda_{1} \RED{e_4} \\
 0 & 0 & 0 & -\frac{\lambda_{3} \RED{c_2}}{\sqrt{10}} & \sqrt{\frac{3}{5}} \alpha  \lambda_{2} \RED{f_4} & m_{351'}-\frac{\lambda_{8} \RED{v}}{2 \sqrt{30}}-\frac{\lambda_{8} \RED{w}}{6 \sqrt{2}}\\
\end{smallmatrix}\right)$},\\
M_{22}&=\scalebox{0.65}{$
\left(\begin{smallmatrix}
m_{351'}+\frac{\sqrt{3}}{4\sqrt{10}} \lambda_{8} \RED{v}-\frac{5 \lambda_{8} \RED{w}}{12 \sqrt{2}} & 0 & 0 & \frac{1}{2} \sqrt{3} \alpha  \lambda_{2} \RED{f_4} & 0 & \frac{1}{2} \sqrt{15} \beta
   \lambda_{2} \RED{f_4} & 0 \\
   0 & m_{351'}-\frac{\lambda_{8} \RED{v}}{2 \sqrt{30}}-\frac{\lambda_{8} \RED{w}}{6 \sqrt{2}} & \sqrt{5} \beta  \lambda_{2} \RED{f_4} & 0 & 0 & 0 & 0 \\
   0 & \sqrt{5} \beta  \lambda_{1} \RED{e_4} & m_{351'}+\frac{\lambda_{8} \RED{v}}{2 \sqrt{30}}-\frac{\lambda_{8} \RED{w}}{6 \sqrt{2}} & 0 & 0 & 0 & 2 \sqrt{10} \lambda_{1} \RED{e_4} \\
   \frac{1}{2} \sqrt{3} \alpha  \lambda_{1} \RED{e_4} & 0 & 0 & m_{351'}-\frac{\sqrt{3}}{4\sqrt{10}}\lambda_{8} \RED{v}+\frac{\lambda_{8} \RED{w}}{12
   \sqrt{2}} & 0 & 0 & 0 \\
   0 & 0 & 0 & 0 & m_{351'}-\frac{7 \lambda_{8} \RED{v}}{4 \sqrt{30}}+\frac{\lambda_{8} \RED{w}}{12 \sqrt{2}} & 0 & 0 \\
   \frac{1}{2} \sqrt{15} \beta  \lambda_{1} \RED{e_4} & 0 & 0 & 0 & 0 & m_{351'}-\frac{\sqrt{3}}{4\sqrt{10}}\lambda_{8} \RED{v}+\frac{\lambda_{8} \RED{w}}{12 \sqrt{2}} & 0 \\
   0 & 0 & 2 \sqrt{10} \lambda_{2} \RED{f_4} & 0 & 0 & 0 & m_{351'}-\frac{\lambda_{8} \RED{v}}{2 \sqrt{30}}-\frac{\lambda_{8} \RED{w}}{6 \sqrt{2}} \\
\end{smallmatrix}\right)$},
\end{align}
and the off-diagonal blocks defined by
\begin{align}
M_{12}&=\scalebox{0.9}{$
\left(\begin{smallmatrix}
-\frac{\lambda_{8} \RED{f_3}}{2 \sqrt{6}} & 0 & 0 & \frac{\alpha  \lambda_{8} \RED{e_4}}{24 \sqrt{2}} & -\frac{\lambda_{8} \RED{f_1}}{2 \sqrt{6}} & \frac{\sqrt{5}\beta \lambda_{8} \RED{e_4}}{24\sqrt{2}} & 0 \\
0 & 0 & 0 & 2 \sqrt{\frac{2}{5}} \lambda_{3} \RED{c_2} & 0 & 0 & 0 \\
0 & 0 & 0 & 0 & -\sqrt{2} \lambda_{4} \RED{d_2} & 0 & 0 \\
0 & -\sqrt{\frac{3}{2}} \lambda_{4} \RED{d_2} & 0 & 0 & 0 & 0 & 0 \\
0 & 0 & 0 & 0 & 0 & 0 & 0 \\
0 & 0 & 0 & 0 & 0 & 0 & 0 \\
\end{smallmatrix}\right)$},\\
M_{21}&=\scalebox{0.9}{$
\left(\begin{smallmatrix}
 -\frac{\lambda_{8} \RED{e_3}}{2 \sqrt{6}} & 0 & 0 & 0 & 0 & 0\\
 0 & 0 & 0 & -\sqrt{\frac{3}{2}} \lambda_{3} \RED{c_2} & 0 & 0\\
 0 & 0 & 0 & 0 & 0 & 0\\
 \frac{\alpha  \lambda_{8} \RED{f_4}}{24 \sqrt{2}} & 2 \sqrt{\frac{2}{5}} \lambda_{4} \RED{d_2} & 0 & 0 & 0 & 0\\
 -\frac{\lambda_{8} \RED{e_1}}{2 \sqrt{6}} & 0 & -\sqrt{2} \lambda_{3} \RED{c_2} & 0 & 0 & 0\\
 \frac{\sqrt{5}\beta  \lambda_{8} \RED{f_4}}{24\sqrt{2}} & 0 & 0 & 0 & 0 & 0\\
 0 & 0 & 0 & 0 & 0 & 0\\
\end{smallmatrix}\right) $}.
\end{align}

The separation of the matrix $\mathcal{M}$ into $4$ blocks is arbitrary and is used above merely as a simple way of presenting a large matrix. The matrix $\mathcal{M}_{\textrm{doublets}}$ is obtained out of $\mathcal{M}$ by removing the last row and column, and taking $\alpha=-3$, $\beta=-\sqrt{3}$, while the matrix $\mathcal{M}_{\textrm{triplets}}$ is obtained by taking $\alpha=\beta=2$. Note that the matrices were already simplified by taking the vacuum ansatz of vanishing VEVs
\begin{align}
c_1=d_1=e_2=f_2=e_5=f_5=u_1=u_2=y&=0,
\end{align}
\noindent
while the rest of the vacuum solution was not plugged-in. Notice also that the coefficients $\alpha$ and $\beta$ are located as factors in front of $e_4$ or $f_4$; this is expected, since $e_4$ and $f_4$ are the $\SU(5)$-breaking VEVs, so they control the difference between the doublets and triplets. Also, the coefficients $\alpha$ are $-3$ and $2$ for the doublets, respectively, which come from the VEV $\langle 24\rangle$ in the terms $5\cdot\langle 24\rangle\cdot\bar{5}$ of $\SU(5)$. The coefficients $\beta$ are $-\sqrt{3}$ and $2$ for the doublets and triplets respectively, which are the Clebsch-Gordan coefficients coming from the terms $5\cdot\langle 24\rangle\cdot \overline{45}$ or $\bar{5}\cdot\langle 24\rangle\cdot 45$ of $\SU(5)$. One can check that this is indeed the case by locating the position of $\alpha$ and $\beta$ coefficients in $\mathcal{M}$ and checking, to which states in Table~\ref{table:labels-doublets-triplets} these matrix entries correspond to.

The procedure for fine-tuning is now similar to the one attempted in the $351'+\overline{351'}+27+\overline{27}$ case in~\cite{Bajc:2013qra}. Once the vacuum solution is plugged-in, the determinants of the two matrices become zero:
\begin{align}
\det(\mathcal{M}_{\textrm{doublets}})\Big|_{\textrm{vacuum}}&=0,\\
\det(\mathcal{M}_{\textrm{triplets}})\Big|_{\textrm{vacuum}}&=0.
\end{align}
The massless doublet-antidoublet and triplet-antitriplet pairs are simply the unphysical would-be Goldstone bosons from the breaking of $\EE$. They correspond to the broken generators with the same quantum numbers, which can be found in $\SO(10)$ language in the $16$ and $\overline{16}$ parts of the adjoint $78$. The doublet-triplet splitting condition then requires another doublet-antidoublet pair to be massless, which imposes the following condition:
\begin{align}
\mathrm{Cond}(\mathcal{M})&:=\mathrm{pmin}_1 \mathcal{M},
\end{align}
where $\mathrm{pmin}_i$ denotes the $i$-th principal minor of rank $1$ (the subdeterminant when the $i$-th row and column are removed). In our specific case, we chose to remove the first row and column, which correspond to the fields $D_0$ and $\overline{D}_0$. We chose these due to simplicity, but note that this choice is valid only if the Goldstone modes of the given vacuum have nonzero $D_0$ and $\overline{D}_0$ components. This can be checked later on via equations~\eqref{equation:Goldstone1} and \eqref{equation:Goldstone2}. The logic behind the whole method is explained in Appendix~\ref{section:simplified-DT}.

Given the notation above, DT-splitting can be performed by a fine-tuning, so that
\begin{align}
\mathrm{Cond}(\mathcal{M}_{\textrm{doublets}})&=0,\label{equation:Cond-D}\\
\mathrm{Cond}(\mathcal{M}_{\textrm{triplets}})&\neq 0.\label{equation:Cond-T}
\end{align}
\noindent
The above conditions can in principle be computed analytically, but they are too complicated to be of any practical use. The viability of the splitting conditions can be shown numerically though, as well as with some careful considerations. The most convenient way to perform the DT splitting is to make use of the dimensionless parameters $\lambda_5$ and $\lambda_6$ from the superpotential (see~\eqref{equation:superpotential-breaking-sector}). Since the invariants $27^3$ and $\overline{27}^3$ have no all-singlet terms, $\lambda_5$ and $\lambda_6$ are not involved in the computation of the vacuum. Moreover, the matrix $\mathcal{M}$ contains only a single entry with the parameter $\lambda_5$ and a single entry with the parameter $\lambda_6$. The condition in equations~\eqref{equation:Cond-D} and \eqref{equation:Cond-T} can be written as
\begin{align}
     K_1-K_2\;\lambda_5\,\lambda_6&=0,\label{equation:Cond-D2}\\
     K'_1-K'_2\;\lambda_5\,\lambda_6&\neq 0,\label{equation:Cond-T2}
     \end{align}
where $K_1,K_2,K'_1,K'_2$ depend only on the other parameters in the Lagrangian ($m_{351'}$, $m_{78}$, $m_{27}$, $\lambda_1$, $\lambda_2$, $\lambda_3$, $\lambda_4$, $\lambda_7$, $\lambda_8$) and the vacuum itself, all of which can be computed independently from the parameters $\lambda_5$ and $\lambda_6$. The analytic forms of $K$s are not very illuminating, but they can easily be evaluated numerically for any values of the parameters $m_{351'},m_{78},m_{27},\lambda_1,\lambda_2,\lambda_3,\lambda_4,\lambda_7,\lambda_8$. One can then take
\begin{align}
\lambda_5&=\frac{K_1}{K_2\,\lambda_6},
\end{align}
with $\lambda_6$ arbitrary, and get an extra massless doublet mode. Substituting into condition~\eqref{equation:Cond-T2}, we can numerically check that indeed
\begin{align}
K'_1 K_2&\neq K'_2 K_1.
\end{align}

We therefore conclude that the addition of the $78$ in the Higgs sector, both with the new vacuum and enlarged matrices $\mathcal{M}_{\textrm{doublets}}$ and $\mathcal{M}_{\textrm{triplets}}$, now allows for a DT splitting in the model. Without the $78$, a similar attempt at fine-tuning is not possible, since in that case we get $K_2=K'_2=0$ after inserting the vacuum and fine-tuning in $\lambda_5$ or $\lambda_6$ is not possible. Note that this tree-level fine-tuning is stable under quantum corrections due to the non-renormalization theorem for the superpotential.

The linear combinations of $D$s and $\bar{D}$s, which correspond to the MSSM Higgses $H_u$ and $H_d$, can be found by computing the new left- and right- null-eigenvectors of the fine-tuned matrix $\mathcal{M}_{\textrm{doublets}}$. In practice, the
physical MSSM Higgses can most easily be extracted by computing the (left and right) null-eigenspace of the fine-tuned matrix $\mathcal{M}_{\textrm{doublets}}$, which is $2$-dimensional. Given any basis of the null-eigenspace, the Higgs will always be orthogonal to the would-be Goldstone boson. The would-be Goldstone itself can be easily identified by the fact that it has no component in the directions $D_i$ (or $\overline{D}_i$) for $i=1,3,4,5,7,8$. This absence of some doublets in the Goldstone can be deduced from the mass matrix in equation~\eqref{equation:DT-matrix}, but we also confirmed this by noting that the Goldstone components $\phi_i$ are the ones which have couplings of their derivatives to the the gauge field through the following type of expression (originating from the kinetic terms of scalar fields):
\begin{align}
-ig\,A_{\mu}^{a}\;(\partial^\mu \phi^\dagger_i) \;(\hat{t}^a \langle\phi\rangle)^i.\label{equation:would-be-Goldstone}
\end{align}
Choosing $a$ to be the doublet/antidoublet broken generators and using our vacuum,  we identify the doublet components in $\phi^\dagger_i$ to which there is no coupling as indeed those listed above. Explicit computation identifies that the prevailing cause of some components not being present in the would-be Goldstone mode is our ansatz of vanishing VEVs. More precisely, without the vacuum ansatz, the terms in expression~\eqref{equation:would-be-Goldstone} can be schematically written as

\begin{align}
\partial D_{G}^\ast&\propto
+\frac{3 \sqrt{5} \RED{v}+5 \sqrt{3} \RED{w}-\sqrt{30} \RED{y}}{10 \sqrt{2}} \;\partial D_{0}^\ast
-\frac{\RED{d_2}}{2} \;\partial D_{2}^\ast
-\frac{\RED{d_1}}{2} \;\partial D_{3}^\ast
+\frac{\RED{e_5}}{4} \sqrt{\frac{3}{2}} \;\partial D_{4}^\ast\nonumber\\
&\qquad +\frac{\sqrt{5} \RED{f_2}}{4} \;\partial D_{5}^\ast
+\frac{\RED{f_3}}{\sqrt{2}} \;\partial D_{6}^\ast
+\frac{\sqrt{3} \RED{f_2}}{4} \partial D_{7}^\ast
+\frac{\RED{e_5}}{4} \sqrt{\frac{5}{2}} \;\partial D_{8}^\ast\nonumber\\
&\qquad -\frac{\RED{e_4}}{4} \sqrt{\frac{3}{2}} \;\partial D_{9}^\ast
+\frac{\RED{f_1}}{\sqrt{2}} \;\partial D_{10}^\ast
-\frac{\RED{e_4}}{4} \sqrt{\frac{5}{2}} \;\partial D_{11}^\ast,\label{equation:Goldstone1}\\
\partial \overline{D}_{G}{}^\ast&\propto
-\frac{3 \sqrt{5} \RED{v}+5 \sqrt{3} \RED{w}-\sqrt{30} \RED{y}}{10 \sqrt{2}} \;\partial \overline{D}_{0}^\ast
+\frac{\RED{c_2}}{2} \;\partial \overline{D}_{2}^\ast
+\frac{\RED{c_1}}{2} \partial \overline{D}_{3}^\ast
-\frac{\RED{f_5}}{4} \sqrt{\frac{3}{2}} \;\partial \overline{D}_{4}^\ast\nonumber\\
&\qquad -\frac{\sqrt{5} \RED{e_2}}{4} \;\partial \overline{D}_{5}^\ast
-\frac{\RED{e_3}}{\sqrt{2}} \;\partial \overline{D}_{6}^\ast
-\frac{\sqrt{3} \RED{e_2}}{4} \;\partial \overline{D}_{7}^\ast
-\frac{\RED{f_5}}{4} \sqrt{\frac{5}{2}} \;\partial \overline{D}_{8}^\ast\nonumber\\
&\qquad +\frac{\RED{f_4}}{4} \sqrt{\frac{3}{2}} \;\partial \overline{D}_{9}^\ast
-\frac{\RED{e_1}}{\sqrt{2}} \;\partial \overline{D}_{10}^\ast
+ \frac{\RED{f_4}}{4} \sqrt{\frac{5}{2}} \;\partial \overline{D}_{11}^\ast.\label{equation:Goldstone2}
\end{align}
We see that the absence of components $D_i$ and $\overline{D}_i$ for $i=3,4,5,7,8$ are directly related to the vacuum ansatz with vanishing VEVs. $D_1$ and $\overline{D}_1$, however, are always absent.\footnote{The $27$ contains $2$ singlets, but $3$ doublets, so the singlets cannot be pushed to all the doublets with a $t^a$ generator, and thus one doublet component is missing. The representation $351'$ has $5$ singlets and $8$ doublets, but the projection relation $d_{ijk}\,351'^{jk}=0$ reshuffles the definitions of singlets and doublets, so there is no loss of doublet components.}

The ratios of the various components of the Higgs, and hence the ratios of the EW VEVs of these component, are computed from the null-eigenspaces, while the magnitudes of the VEVs are given by the VEVs $v_u=\langle H_u\rangle $ and $v_d=\langle H_d\rangle$. We have
\begin{align}
v_u^2&=\sum_{i=0}^{11} |\BLUE{v_i}|^2,\\
v_d^2&=\sum_{i=0}^{11} |\BLUE{\bar{v}_i}|^2,
\end{align}
where $v_i=\langle D_i\rangle$ and $\bar{v}_i=\langle \overline{D}_i\rangle$, and additionally the following MSSM relations hold:
\begin{align}
v_u^2+v_d^2&=(246\,\mathrm{GeV})^2,\\
v_u/v_d&=\tan\beta.
\end{align}

\section{Yukawa sector\label{section:Yukawa-sector}}

The Yukawa sector comes from the Yukawa part of the superpotential after inserting the vacuum solution:
\begin{align}
W_{\textrm{Yukawa}}&=27_F^i\;27_F^j\;\big(Y^{ij}_{27}\;\langle 27\rangle + Y_{\overline{351'}}^{ij}\;\langle \overline{351'}\rangle\,\big).
\end{align}

In addition to the GUT scale VEVs in the representations $\overline{351'}$ and $27$, the EW Higgses $H_u$ and $H_d$ also need to be present in both of these representations. The low-energy MSSM Higgses $H_u$ and $H_d$ come from a linear combination of the doublets of type $D\sim(1,2,+1/2)$ and antidoublets of the type $\overline{D}\sim(1,2,-1/2)$, respectively. These states are identified in Table~\ref{table:labels-doublets-triplets}, and their EW VEVs are labeled by $v_i:=\langle D_i\rangle$ and $\bar{v}_i:=\langle \overline{D}_i \rangle$, where $i=0,1,\ldots,11$. As discussed in section~\ref{section:DT-splitting}, we generically expect $v_i\neq 0$ and $\bar{v}_i\neq 0$.

Each generation of fermions is found in a fundamental representation $27$ of $\EE$. The matter content of each generation, subdivided into $\SO(10)$ representations, is the following:
\begin{itemize}
\item The $16$ of $\SO(10)$ contains the SM particles\footnote{We use the standard notation $Q$, $L$, $u^c$, $d^c$ and $e^c$ for the representations $(3,2,+1/6)$, $(1,2,-1/2)$, $(\bar{3},1,-2/3)$, $(\bar{3},1,1/3)$ and $(1,1,1)$, respectively. The lepton doublet contains the electron $e$ and the neutrino $\nu$.} and the right-handed neutrino $\nu^c$.
\item The $10$ of $\SO(10)$ contains a vector-like pair of down-type quarks $(3,1,-1/3)+(\bar{3},1,+1/3)$, as well as a vector-like pair of lepton doublets $(1,2,-1/2)+(1,2,1/2)$. These exotic are labeled $d'$, $d'^c$, $L'$ and $L'^c$, respectively.
\item The $\SO(10)$ singlet $1$ is denoted by $s$ is sterile, and has a role analogous to the right-handed neutrino.
\end{itemize}
Suppressing generation indices, the mass terms are computed to be
\begin{align}
&u^{T}\left(-\BLUE{v_1} \yd+\tfrac{1}{2\sqrt{10}} \BLUE{v_5}\yt-\tfrac{1}{2\sqrt{6}} \BLUE{v_7}\yt\right) u^c\nonumber\\
&+
\begin{pmatrix}
d^{cT} & d'^{cT}\\
\end{pmatrix}
\begin{pmatrix}
 \phantom{-}\BLUE{\bar{v}_2} \yd+\tfrac{1}{2\sqrt{10}} \BLUE{\bar{v}_4}\yt+\tfrac{1}{2\sqrt{6}} \BLUE{\bar{v}_8}\yt  & \phantom{-}\RED{c_2} \yd+\frac{1}{\sqrt{15}}\RED{f_5} \yt  \\
 -\BLUE{\bar{v}_3} \yd-\tfrac{1}{2\sqrt{10}} \BLUE{\bar{v}_9}\yt-\tfrac{1}{2\sqrt{6}} \BLUE{\bar{v}_{11}}\yt  & -\RED{c_1}\yd+\frac{1}{\sqrt{15}}\RED{f_4} \yt \\
\end{pmatrix}\!\!
\begin{pmatrix}
d \\
d'\\
\end{pmatrix} \nonumber\\
&+
\begin{pmatrix}
e^{T} & e'^{T}
\end{pmatrix}
\begin{pmatrix}
 -\BLUE{\bar{v}_2} \yd- \frac{1}{2\sqrt{10}} \BLUE{\bar{v}_4}\yt+ \sqrt{\tfrac{3}{8}} \BLUE{\bar{v}_8} \yt  & \phantom{-}\RED{c_2} \yd- \tfrac{3}{2}\frac{1}{\sqrt{15}} \RED{f_5} \yt  \\
 \phantom{-}\BLUE{\bar{v}_3} \yd+\frac{1}{2\sqrt{10}}\BLUE{\bar{v}_9}\yt-\sqrt{\tfrac{3}{8}}\BLUE{\bar{v}_{11}} \yt  & -\RED{c_1} \yd- \tfrac{3}{2}\frac{1}{\sqrt{15}} \RED{f_4} \yt \\
\end{pmatrix}\!\!
\begin{pmatrix}
e^c \\
e'^c\\
\end{pmatrix} &&\nonumber\\
&+
\begin{pmatrix}
\nu^{T} & \nu'^{T}\\
\end{pmatrix} \!\!
\begin{pmatrix}
 \BLUE{v_1} \yd-\tfrac{1}{2\sqrt{10}}\BLUE{v_5}\yt-\sqrt{\frac{3}{8}} \BLUE{v_7}\yt  & -\tfrac{1}{\sqrt{2}} \BLUE{v_6} \yt & \phantom{-}\RED{c_2} \yd-\tfrac{3}{2}\frac{1}{\sqrt{15}} \RED{f_5} \yt  \\
 -\tfrac{1}{\sqrt{2}} \BLUE{v_{10}} \yt & -\BLUE{v_1} \yd-\sqrt{\frac{2}{5}} \BLUE{v_5} \yt  & -\RED{c_1} \yd-\tfrac{3}{2}\frac{1}{\sqrt{15}} \RED{f_4} \yt \\
\end{pmatrix} \!\!
\begin{pmatrix}
\nu^c \\
s \\
\nu'^c
\end{pmatrix} \nonumber\\
&+\frac{1}{2}
\begin{pmatrix}
\nu^{cT} & s^T & \nu'^{cT}\\
\end{pmatrix}\!
\begin{pmatrix}
 \phantom{\tfrac{1}{\sqrt{2}}}\RED{f_1} \yt  & \tfrac{1}{\sqrt{2}}\RED{f_2} \yt &-\BLUE{\bar{v}_3} \yd+ \sqrt{\frac{2}{5}} \BLUE{\bar{v}_9} \yt \\
 \tfrac{1}{\sqrt{2}}\RED{f_2} \yt & \phantom{\tfrac{1}{\sqrt{2}}}\RED{f_3} \yt  &\phantom{-}\BLUE{\bar{v}_2} \yd-\sqrt{\frac{2}{5}} \BLUE{\bar{v}_4} \yt  \\
 -\BLUE{\bar{v}_3} \yd+\sqrt{\frac{2}{5}} \BLUE{\bar{v}_9} \yt & \BLUE{\bar{v}_2} \yd-\sqrt{\frac{2}{5}} \BLUE{\bar{v}_4} \yt  & 0\\
\end{pmatrix}\!\!
\begin{pmatrix}
\nu^c \\
s \\
\nu'^c\\
\end{pmatrix}\nonumber\\
&+\frac{1}{2}
\begin{pmatrix}
\nu^T & \nu'^T
\end{pmatrix}
\begin{pmatrix}
\Delta_1 Y_{\overline{351'}} & \tfrac{1}{\sqrt{2}}\Delta_2 Y_{\overline{351'}}\\
\tfrac{1}{\sqrt{2}}\Delta_2 Y_{\overline{351'}}& \Delta_3 Y_{\overline{351'}}\\
\end{pmatrix}
\begin{pmatrix}
\nu \\
\nu' \\
\end{pmatrix}\!.\label{equation:Yukawa-terms-full}
\end{align}

Notice the different Clebsch-Gordan coefficients in front of $\bar v_i$ ($i=2,3,4,9$) that come from the couplings with the
$\bar 5$ of $\SU(5)$ and in front of $\bar v_{8,11}$ that come from the $\overline{45}$, while $c_{1,2}$ originate from a $\SU(5)$
singlet and $f_{4,5}$ from a $\SU(5)$ adjoint $24$.

The $\Delta_i$ correspond to VEVs induced in the electrically neutral component of weak triplets of type $(1,3,+1)$, while $\overline{\Delta}_i$ correspond to VEVs induced in weak triplets $(1,3,-1)$. The $\Delta_i$ and $\overline{\Delta}_i$ are found only in the representations $351'$ and $\overline{351'}$, as shown in the definitions of Table~\ref{table:weak-triplets}. The mass matrix $M_{\Delta}$ has contributions from the terms $351'\cdot\overline{351'}$ and $351'\cdot\langle 78\rangle\cdot\overline{351'}$. Its explicit form is computed to be
    \begin{align}
    M_\Delta&=\scalebox{0.55}{$
    \begin{pmatrix}
     m_{351'}-\lambda _8 \left(\frac{\RED{w}}{6 \sqrt{2}}-\frac{1}{2} \sqrt{\frac{3}{10}} \RED{v}+\frac{\RED{y}}{2 \sqrt{5}}\right)
   & \frac{\RED{u_1} \lambda _8}{2 \sqrt{3}} & 0 & \phantom{-}6 \lambda _1 \RED{e_1} \\
    \frac{\lambda _8}{2 \sqrt{3}}\RED{u_2} & m_{351'}+\lambda_8 \left(\frac{\RED{w}}{12 \sqrt{2}}+\frac{\RED{v}}{4 \sqrt{30}}-\frac{\RED{y}}{2 \sqrt{5}}\right)& \frac{\lambda_8}{2 \sqrt{3}}\RED{u_1} & -6 \lambda _1 \RED{e_2}
   \\
    0 & \frac{\lambda _8}{2 \sqrt{3}}\RED{u_2} & m_{351'}+ \lambda _8\left(\frac{\RED{w}}{3 \sqrt{2}}-\frac{\RED{v}}{\sqrt{30}}-\frac{\RED{y}}{2 \sqrt{5}}\right) & \phantom{-}6 \lambda _1\RED{e_3} \\
    6 \lambda _2\RED{f_1} & -6 \lambda _2\RED{f_2} & 6 \lambda _2 \RED{f_3} & m_{351'}+ \lambda _8 \left(\frac{\RED{w}}{3 \sqrt{2}}+\frac{\RED{v}}{\sqrt{30}}+\frac{\RED{y}}{2 \sqrt{5}}\right)
    \end{pmatrix}
    $}.
    \end{align}
Integrating out the heavy weak triplets and inserting the $F$-term ansatz of vanishing VEVs, we get
    \begin{align}
    \begin{pmatrix}
    \Delta_1\\ \Delta_2\\\Delta_3\\\Delta_4\\
    \end{pmatrix}&=
    \scalebox{0.7}{$
    \left(\begin{smallmatrix}
     m_{351'}-\lambda _8 \left(\frac{\RED{w}}{6 \sqrt{2}}-\frac{1}{2} \sqrt{\frac{3}{10}} \RED{v}\right)
   & 0 & 0 & 6 \lambda _1 \RED{e_1} \\
    0 & m_{351'}+\lambda_8 \left(\frac{\RED{w}}{12 \sqrt{2}}+\frac{\RED{v}}{4 \sqrt{30}}\right)& 0 & 0
   \\
    0 & 0 & m_{351'}+ \lambda _8\left(\frac{\RED{w}}{3 \sqrt{2}}-\frac{\RED{v}}{\sqrt{30}}\right) & 6 \lambda _1\RED{e_3} \\
    6 \lambda _2\RED{f_1} & 0 & 6 \lambda _2 \RED{f_3} & m_{351'}+ \lambda _8 \left(\frac{\RED{w}}{3 \sqrt{2}}+\frac{\RED{v}}{\sqrt{30}}\right)
    \end{smallmatrix}\right)$}^{-1}
    \begin{pmatrix}
    \lambda_4 \BLUE{v_3}^2\\ \lambda_4 \sqrt{2}\BLUE{v_2} \BLUE{v_3}\\ \lambda_4 \BLUE{v_2}^2\\ \lambda_3 \BLUE{v_1}^2\\
    \end{pmatrix}.
    \end{align}

After integrating out the heavy vector-like states from equation~\eqref{equation:Yukawa-terms-full}, and using the ansatz $c_1=f_5=0$, we get the matrices for the low energy states:

\begin{align}
M_U&=-\BLUE{v_1} \yd+\left(\tfrac{1}{2\sqrt{10}} \BLUE{v_5}-\tfrac{1}{2\sqrt{6}} \BLUE{v_7}\right)\yt,\label{equation:low-energy-masses-first}\\
M_D^T&=\scalebox{0.9}{$ \left(1+(9/4)XX^\dagger\right)^{-1/2} \left(\left(\BLUE{\bar{v}_2}-\tfrac{3}{2}\BLUE{\bar{v}_3}X
\right)\yd+\left(\tfrac{1}{2\sqrt{10}}(\BLUE{\bar{v}_4}-\tfrac{3}{2}\BLUE{\bar{v}_9}X)+\tfrac{1}{2\sqrt{6}}(\BLUE{\bar{v}_8}-\tfrac{3}{2}\BLUE{\bar{v}_{11}}X)\right)\yt\right) $},\\
M_E&=\scalebox{0.9}{$ \left(1+XX^\dagger\right)^{-1/2} \left(\left(-\BLUE{\bar{v}_2}-\BLUE{\bar{v}_3}X
\right)\yd+\left(-\tfrac{1}{2\sqrt{10}}(\BLUE{\bar{v}_4}+\BLUE{\bar{v}_9}X)+\sqrt{\tfrac{3}{8}}(\BLUE{\bar{v}_8}+\BLUE{\bar{v}_{11}}X)\right)\yt\right) $},\label{equation:low-energy-masses-third}\\
M_N&=-\left(1+XX^\dagger\right)^{-1/2}\nonumber\\
    &\Bigg(\scalebox{0.95}{$ \bigg(-\frac{1}{\sqrt{10}}\frac{\BLUE{v_1} \BLUE{v_5}}{\RED{f_1}}-\sqrt{\frac{3}{2}}\frac{\BLUE{v_1} \BLUE{v_7}}{\RED{f_1}}+
    \frac{1}{\sqrt{3}}\frac{\BLUE{v_5} \BLUE{v_{10}}}{\RED{f_1}}\frac{\RED{c_2}}{\RED{f_4}}+\sqrt{5}\,\frac{\BLUE{v_7} \BLUE{v_{10}}}{\RED{f_1}}\frac{\RED{c_2}}{\RED{f_4}}+
    \frac{4}{\sqrt{3}}\frac{\BLUE{v_5} \BLUE{v_6}}{\RED{f_3}}\frac{\RED{c_2}}{\RED{f_4}}-2\sqrt{\frac{10}{3}}\,\Delta_2\,\frac{\RED{c_2}}{\RED{f_4}}\bigg) $}\;\yd\nonumber\\
    &\quad +\bigg(\frac{1}{40}\frac{\BLUE{v_5}^2}{\RED{f_1}}+\sqrt{\frac{3}{80}}\frac{\BLUE{v_7} \BLUE{v_5}}{\RED{f_1}}+\frac{3}{8}\frac{\BLUE{v_7}^2}{\RED{f_1}}+ \frac{1}{2}\frac{\BLUE{v_6}^2}{\RED{f_3}}-\Delta_1\bigg)\;\yt\nonumber\\
    &\quad + \scalebox{0.95}{ $\bigg(\frac{\BLUE{v_1}^2}{\RED{f_1}}-2 \sqrt{\frac{10}{3}} \frac{\BLUE{v_1} \BLUE{v_{10}}}{\RED{f_1}}\frac{\RED{c_2}}{\RED{f_4}}
    +\frac{10}{3}\frac{\BLUE{v_{10}}^2}{\RED{f_1}}\frac{\RED{c_2}^2}{\RED{f_4}^2}
    +2 \sqrt{\frac{10}{3}}\frac{\BLUE{v_1} \BLUE{v_6}}{\RED{f_3}}\frac{\RED{c_2}}{\RED{f_4}}
    +\frac{8}{3}\frac{\BLUE{v_5}^2}{\RED{f_3}}\frac{\RED{c_2}^2}{\RED{f_4}^2}-\frac{20}{3}\,\Delta_3\,\frac{\RED{c_2}^2}{\RED{f_4}^2}\bigg) $}\;\yz\nonumber\\
    &\quad +\bigg(\frac{8\sqrt{10}}{3}\frac{\BLUE{v_1} \BLUE{v_5}}{\RED{f_3}}\frac{\RED{c_2}^2}{\RED{f_4}^2}\bigg)\;\yd\yt^{-1}\yd\yt^{-1}\yd\nonumber\\
    &\quad +\bigg(\frac{20}{3}\frac{\BLUE{v_1}^2}{\RED{f_3}}\frac{\RED{c_2}^2}{\RED{f_4}^2}\bigg)\;\yd\yt^{-1}\yd\yt^{-1}\yd\yt^{-1}\yd\Bigg)\nonumber\\
    &\left(1+X^\ast X^T\right)^{-1/2}\label{equation:low-energy-masses-last},
\end{align}
where
\begin{align}
X&=-2\sqrt{\tfrac{5}{3}}\,\frac{\RED{c_2}}{\RED{f_4}}\,\yd\,\yt^{-1}.\label{equation:X0-definition}
\end{align}

Notice that the main factor in the expressions is a linear combination of the following matrices:
\begin{align}
\yd,\quad \yt,\quad \yd\yt^{-1}\yd, \quad \yd\yt^{-1}\yd\yt^{-1}\yd,\quad \yd\yt^{-1}\yd\yt^{-1}\yd\yt^{-1}\yd,
\end{align}
which are all symmetric, since $\yd$ and $\yt$ are symmetric, as are $M_U$ and $M_N$.
The matrices $M_D$ and $M_E$ are not symmetric, though, due to the projection factor onto the light families.

As usual with vector-like states, the expressions for the low energy masses in equations~\eqref{equation:low-energy-masses-first}--\eqref{equation:low-energy-masses-last} are nonlinear, which complicates the analysis of the masses. We comment more on the low-energy part of the Yukawa sector in section~\ref{section:numeric-fit}, in which we also do a numeric fit in the 2-generation case.

\section{Numeric fit of the Yukawa sector\label{section:numeric-fit}}
The presented model has $3$ masses and $8$ dimensionless parameters $\lambda_i$ in the breaking sector, as well as two symmetric $3\times 3$ Yukawa matrices in the Yukawa sector, which is easily seen from the superpotential in equation~\eqref{equation:W-full}. Note also that the product $\lambda_5\lambda_6$ is fixed by the fine-tuning of the EW Higgs mass. Also, a rotation in family-space can bring one of the Yukawa matrices to be diagonal. We shall limit ourselves to the case where all parameters of the Lagrangian are real. The independent number of real parameters in the breaking sector is thus $3+8-1=10$, while the $3$-family Yukawa sector has $6+3=9$ real parameters. Since the number of independent real parameters is $19$, while there are only $17$ real numbers to be measured ($3$ masses in the up, down, charged lepton sector each, the two differences of masses-squared in the neutrino sector, as well as $3$ angles in each of the CKM and PMNS matrices; we neglect the CP-phases), the general expectation is that a fit is possible to perform. There may exist, however, non-obvious mass relations concealed due to the complexity of the low energy expressions in equations~\eqref{equation:low-energy-masses-first}--\eqref{equation:low-energy-masses-last}, which are not respected by the experimental values; these fears can be alleviated by finding points in parameter space, which give a good fit for to the masses and mixing angles.

In this section, we perform a fit in the simplified case of $2$ families. Here, there are again $10$ real parameters in the breaking sector, while the Yukawa sector has $3+2=5$ real parameters. The fit is performed for the results of the quark masses $m_{t}$, $m_{c}$, $m_{b}$, $m_{s}$, the charged lepton masses $m_{\tau}$ and $m_{\mu}$, the difference of the squared neutrino masses $m_{\nu_3}^2-m_{\nu_2}^2$, and the mixing angles $\theta_{cb}$ and $\theta_{23}$ in the CKM and PMNS matrices, respectively.

The most convenient way to perform the fit is to take some of the GUT scale VEVs as the parameters in the fit, instead of the initial parameters in the Lagrangian. This is advantageous since the equations of motion are linear in the Lagrangian parameters. Taking the ansatz $c_1=d_1=e_2=f_2=e_5=f_5=u_1=u_2=y=0$, we need $12$ parameters; only $10$ are independent, while $2$ are determined through the $D$-terms. It is convenient, for example, to take the following quantities as independent parameters:
\begin{align}
m_{351'},\quad c_2,\quad d_2,\quad f_1,\quad f_3,\quad f_4,\quad e_4,\quad v,\quad w,\quad \lambda_5.\label{equation:fit-parameters}
\end{align}

We can then use the equations of motion to determine the remaining initial parameters and VEVs. The $D$-terms are solved by taking
    \begin{align}
    \label{d2}
    e_1&= \pm\sqrt{|\RED{d_2}|^2 - |\RED{c_2}|^2 + 2 |\RED{f_1}|^2}/\sqrt{2},\\
    \label{e4}
    e_3&= \pm\sqrt{|\RED{e_4}|^2 - |\RED{f_4}|^2 + 2 |\RED{f_3}|^2}/\sqrt{2}.
    \end{align}
\noindent
Since we want $e_1$ and $e_3$ to be real numbers, the arguments in the square root need to be positive, which limits the space of parameters in equation~\eqref{equation:fit-parameters}. Alternatively, we could also take the $6$ independent parameters, for example\footnote{In both parametrizations of encoding the $D$-terms, we wrote them so that we retained control to make $f_1$ and $f_3$ potentially small; this will be important for the neutrino sector, as described later in this section.}, to be the VEVs $d_2$, $f_1$, $e_3$, $f_4$ (now with no restriction) and two angles $\theta_{1},\theta_2\in (0,2\pi)$, with the remaining quantities computed as
\begin{align}
f_1&=\sin\theta_1\;\sqrt{|\RED{c_2}|^2 + 2 |\RED{e_1}|^2}/\sqrt{2},\\
d_2&=\cos\theta_1\;\sqrt{|\RED{c_2}|^2 + 2 |\RED{e_1}|^2},\\
f_3&=\sin\theta_2\;\sqrt{|\RED{f_4}|^2 + 2 |\RED{e_3}|^2}/\sqrt{2},\\
\label{ee4}
e_4&=\cos\theta_2\;\sqrt{|\RED{f_4}|^2 + 2 |\RED{e_3}|^2}.
\end{align}

The $F$-terms then yield

    \begin{align}
    m_{27}&= \frac{m_{351'}}{2 \RED{w} \left(\sqrt{15} \RED{w}-\RED{v}\right) \RED{c_2} \RED{d_2} \left(4 \RED{e_3} \RED{f_3}+\RED{e_4} \RED{f_4}\right)} \left(2 \RED{v}(\RED{w}+\sqrt{15} \RED{v}) \left(\RED{e_4} \RED{f_4}-2 \RED{e_3} \RED{f_3}\right){}^2\right.\nonumber\\
    &\left. +\RED{e_1} \RED{f_1}
    (\RED{v}-\sqrt{15} \RED{w}) \left(2 \RED{e_3}\RED{f_3}(3 \sqrt{15} \RED{v}-5 \RED{w})-\RED{e_4} \RED{f_4}(3 \sqrt{15} \RED{v}+7 \RED{w})\right)\right),\\
    m_{78}&= \frac{\sqrt{15} m_{351'} \left(\RED{e_4} \RED{f_4}-2 \RED{e_3} \RED{f_3}\right){}^2}{2 \RED{w}
   \left(\sqrt{15} \RED{w}-\RED{v}\right) \left(4 \RED{e_3} \RED{f_3}+\RED{e_4} \RED{f_4}\right)},\\
    \lambda _1&=-\frac{\RED{f_3} \RED{f_4} m_{351'}}{\RED{e_4} (\RED{e_4} \RED{f_4}+4 \RED{e_3} \RED{f_3})},\\
    \label{lambda2}
    \lambda _2&=-\frac{\RED{e_3} \RED{e_4} m_{351'}}{\RED{f_4} (\RED{e_4} \RED{f_4}+4 \RED{e_3} \RED{f_3})},\\
    \lambda _3&=-\frac{\RED{e_1} m_{351'} \left(\sqrt{15} \RED{v} \left(\RED{e_4} \RED{f_4}-2 \RED{e_3} \RED{f_3}\right)+3 \RED{w} \left(2 \RED{e_3} \RED{f_3}+\RED{e_4} \RED{f_4}\right)\right)}{2 \RED{w} \RED{c_2}^2 \left(4 \RED{e_3} \RED{f_3}+\RED{e_4} \RED{f_4}\right)},\\
    \lambda _4&=-\frac{\RED{f_1} m_{351'} \left(\sqrt{15} \RED{v} \left(\RED{e_4} \RED{f_4}-2 \RED{e_3} \RED{f_3}\right)+3 \RED{w} \left(2 \RED{e_3} \RED{f_3}+\RED{e_4} \RED{f_4}\right)\right)}{2 \RED{w} \RED{d_2}^2 \left(4 \RED{e_3} \RED{f_3}+\RED{e_4} \RED{f_4}\right)},\\
    \lambda _7&= \frac{3 \sqrt{2} m_{351'} \left(2 \RED{e_3} \RED{f_3}-\RED{e_4} \RED{f_4}\right) \left(\left(\RED{v}-\sqrt{15} \RED{w}\right) \RED{e_1} \RED{f_1}
    +2\RED{v}\left(2 \RED{e_3} \RED{f_3}-\RED{e_4} \RED{f_4}\right)\right)}{\RED{w} \left(\sqrt{15} \RED{w}-\RED{v}\right) \RED{c_2} \RED{d_2} \left(4 \RED{e_3} \RED{f_3}+\RED{e_4} \RED{f_4}\right)},\\
    \lambda _8&= \frac{3 \sqrt{2} m_{351'} \left(2 \RED{e_3} \RED{f_3}-\RED{e_4} \RED{f_4}\right)}{\RED{w} \left(\RED{e_4} \RED{f_4}+4 \RED{e_3} \RED{f_3}\right)}.
    \end{align}

Using these values, $\lambda_6$ can then be determined by fine-tuning in the doublet mass matrix (see section~\ref{section:DT-splitting}), from which also the EW VEVs $v_i$ and $\bar{v}_i$ are computed. One can then use these to compute the mass matrices from equations~\eqref{equation:low-energy-masses-first}--\eqref{equation:low-energy-masses-last}, and ultimately the masses and the mixing angles. For our numeric fit, we assume:

\begin{itemize}

\item
the simplified case of second and third generation only;

\item
all parameters real;

\item
all errors in measured quantities at the $10\%$ level; we believe that at the present stage such a choice is a good compromise between the realistic case and
the simplicity of the analysis;

\item
the values for the masses and mixing angles at the GUT scale $\sim 10^{16}\,\mathrm{GeV}$ as shown in Table~\ref{table:numeric-masses-and-angles}~\cite{Xing:2007fb}, valid for $\tan{\beta}=10$;

\item
positive signs in eqs. \eqref{d2}-\eqref{e4}.

\end{itemize}

We performed the fit by minimizing the chi-squared function \begin{align}
\chi^2&:=\sum_{i}{\frac{(f_i(x)-y_i)}{\sigma_i^2}},
\end{align}
with respect to the initial parameters $x$. The experimental values are denoted by $y_i$, the values computed using our model are $f_i(x)$, and $\sigma_i$ are the $1$-sigma deviations from the values $y_i$. For our fit, we have $9$ different measured quantities, thus $i=1,\ldots,9$. The errors are taken to be $\sigma_i=0.1\,y_i$, giving
\begin{align}
\chi^2=100\sum_i \left(\frac{f_i(x)-y_i}{y_i}\right)^2.
\end{align}
We define the convenient measure $\hat{\chi}^2=\chi^2/9$, which tells us the average $\sigma^2$ deviation per measured value. Also, we define the pulls $\chi_i=(f_i(x)-y_i)/\sigma_i$, which tell us how many sigma a certain quantity deviates from the measured one.

Due to the large number of parameters, the $\chi^2$ function will have many local minima in the parameter space. We give below two such points in the parameter space, corresponding to the best fits that were found and which we deem sufficiently good (with $\hat{\chi}^2 \lesssim 1$). The results are given in Tables~\ref{table:numeric-independent-parameters}, \ref{table:numeric-Lagrangian-parameters}, \ref{table:numeric-masses-and-angles}; the points in the parameter space are given in terms of the independent parameters best suited to a numeric search (as discussed in this section) in Table~\ref{table:numeric-independent-parameters} and in terms of the original Lagrangian parameters in Table~\ref{table:numeric-Lagrangian-parameters}, while Table~\ref{table:numeric-masses-and-angles} shows the obtained results for the masses and mixing angles.

Notice from Tables~\ref{table:numeric-independent-parameters} and \ref{table:numeric-Lagrangian-parameters} that of the two Yukawa matrices $\yd$ and $\yt$, $\yt$ was chosen to be the diagonal one. Furthermore, the original Lagrangian parameters in Table~\ref{table:numeric-Lagrangian-parameters} are given so that it can be checked they roughly fall into the perturbative regime. A possible
exception could be the value $\lambda_8$ of the first solution.

Note that the given two points are merely the best ones we found. Due to the high dimensionality of the parameter space, we suspect there are likely many more points which give a comparable or a better fit. There are a number of observations that can be made about these points in general, however, by deducing them from the formulae, comparing the two parameter points given in the tables and through experience  obtained by  performing the fit:
\begin{itemize}
\item \textit{The parameter points are not necessarily close to each other, which also holds true for any specific single parameter.} We can see in Table~\ref{table:numeric-independent-parameters} that it is not necessary for any parameter to be at a very specific value to obtain a good fit. The suitable areas of parameter space thus form many disconnected regions, and no specific value can be claimed for any parameter. In this sense, the mass formulae of the theory are not very predictive of the original parameters.
\item \textit{There is no specific mass or mixing angle, where one would consistently be getting tension.} As observed from Table~\ref{table:numeric-masses-and-angles}, while a specific solution might have most tension with observation coming from a single mass or mixing angle, there are other points, which also give a good fit and where this parameter is predicted better. In this sense we cannot claim any tendencies in tensions of the observables.
\item Imagine that in Table~\ref{table:numeric-independent-parameters} or \ref{table:numeric-Lagrangian-parameters} we rescale all the mass parameters with a common factor. Such a rescaling would have no influence on the masses of the quarks and charged leptons, which are controlled only by the EW VEVs, as can be seen from equations~\eqref{equation:low-energy-masses-first}--\eqref{equation:low-energy-masses-third} and equation~\eqref{equation:X0-definition}. Rescaling would influence the neutrino masses, however, due to the seesaw mechanism, confirmed by equation~\eqref{equation:low-energy-masses-last}. In principle, the rescaling factor can always be adjusted, so that the fit of the mass-square difference of the neutrino masses $m_{\nu_3}^2-m_{\nu_2}^2$ is exact (provided this does not spoil the GUT scale, or the upper bounds on neutrino masses). \textit{Thus only neutrinos are actually sensitive to the GUT scale.}
\item The mass parameters are chosen to be at approximately the GUT scale, say at the order of $10^{15-16}\,\mathrm{GeV}$. If this is true for all the mass parameters, there might be a problem with the neutrino masses. It is a well known fact that this GUT scale is a few orders of magnitude too large compared to the seesaw scale for sufficiently large mass differences in the neutrino sector. One can cure this problem by having the spontaneous symmetry breaking occur in multiple stages, with a mass hierarchy between different VEVs and the seesaw scale corresponding to one of the intermediate stages. The gauge coupling unification in such a scenario could then be spoilt by the particles appearing at these intermediate mass scales.

In our case, however, the number of different VEVs is large enough, so that having one or two of the VEVs at a smaller scale does not disturb the breaking pattern. To see this, note that to obtain sufficiently large neutrino masses, it is enough for one of the terms in equation~\eqref{equation:low-energy-masses-last} to be of the proper scale, which can be achieved by simply taking $f_1$ or $f_3$ to be several orders of magnitude smaller than the GUT scale (the seesaw type~I contributions). In the solutions given in Table~\ref{table:numeric-independent-parameters}, we achieved sufficiently high neutrino masses by taking the parameter$f_1$ (and also $v$) to be a few orders of magnitude below the GUT scale. As seen from the gauge boson masses in Table~\eqref{table:gauge-boson-masses}, small $f_1$ and $v$ do not spoil the one-stage breaking scenario; since the intermediate scales are not associated to an intermediate symmetry breaking at that scale, but are instead purely accidental due to a carefully chosen parameter point, we do not expect too many relevant states (apart from the singlets $\nu^c$ with $f_1$ Majorana mass) to be found at the scale $f_1$. We noticed however the appearance of a lighter color triplet-antitriplet pair, possibly due to the similarities between the doublet and triplet mass matrices.
Although this influences the running of the gauge couplings, we will neglect it in view of the (presumably) large threshold uncertainties present anyway.
\item A final comment on the neutrino masses: although only the difference of masses-squared needs to be fitted, one still needs to check that the neutrino masses themselves are $< 1\,\mathrm{eV}$ \cite{Agashe:2014kda}. As seen in Table~\ref{table:numeric-masses-and-angles}, this condition holds true for both of our parameter points.
\end{itemize}

\begin{table}[H]
\caption{Two example points in the parameter space, written in terms of the independent parameters suitable for a search.\label{table:numeric-independent-parameters}}
\begin{longtable*}{lrr}
\toprule
\multicolumn{1}{c}{parameter}&\multicolumn{1}{c}{point 1}&\multicolumn{1}{c}{point 2}\\
\midrule
$m_{351'}\;[\mathrm{GeV}]$&$ 1.17\times 10^{16} $&$ -4.17\times 10^{15} $\\
$c_2\;[\mathrm{GeV}]$&$ 6.68\times 10^{15} $&$ 3.98\times 10^{15} $\\
$d_2\;[\mathrm{GeV}]$&$ -6.78\times 10^{15} $&$ -4.90\times 10^{15} $\\
$f_1\;[\mathrm{GeV}]$&$ 4.12\times 10^{11} $&$ -5.52\times 10^{12} $\\
$f_3\;[\mathrm{GeV}]$&$ -1.84\times 10^{16} $&$ 1.38\times 10^{16} $\\
$f_4\;[\mathrm{GeV}]$&$ 1.61\times 10^{16} $&$ 1.49\times 10^{16} $\\
$e_4\;[\mathrm{GeV}]$&$ 5.27\times 10^{15} $&$ -1.69\times 10^{16} $\\
$v\;[\mathrm{GeV}]$&$ -7.07\times 10^{13} $&$ 8.44\times 10^{14} $\\
$w\;[\mathrm{GeV}]$&$ 6.13\times 10^{15} $&$ -1.78\times 10^{16} $\\
$\lambda_5$&$ -1.58\times 10^{-3} $&$ 1.50\times 10^{-1} $\\
$(Y_{27})_{11}$&$ -0.723 $&$ 1.93 $\\
$(Y_{27})_{12}$&$ 0.703 $&$ -1.19 $\\
$(Y_{27})_{22}$&$ -0.676 $&$ 0.730$\\
$(Y_{\overline{351'}})_{11}$&$ -0.371 $&$ 0.733$\\
$(Y_{\overline{351'}})_{22}$&$ 0.363 $&$ -0.287$\\
\bottomrule
 \end{longtable*}
\end{table}

\begin{table}[H]
\caption{Two example points in the parameter space presented in terms of the original parameters in the superpotential.}
\label{table:numeric-Lagrangian-parameters}
\begin{longtable*}{lrr}
\toprule
\multicolumn{1}{c}{parameter}&\multicolumn{1}{c}{point 1}&\multicolumn{1}{c}{point 2}\\
\midrule
$m_{351'}\;[\mathrm{GeV}]$&$ 1.17\times 10^{16} $&$ -4.17\times 10^{15} $\\
$m_{27}\;[\mathrm{GeV}]$&$ -2.92\times 10^{14} $&$ -1.64\times 10^{15} $\\
$m_{78}\;[\mathrm{GeV}]$&$ -6.19\times 10^{16} $&$ -5.04\times 10^{15} $\\
$\lambda_1$&$ -6.49\times 10^{-1} $&$ -8.91\times 10^{-2} $\\
$\lambda_2$&$ 5.66\times 10^{-2} $&$ -1.24\times 10^{-1} $\\
$\lambda_3$&$ -1.5\times 10^{-1} $&$ 2.78\times 10^{-1} $\\
$\lambda_4$&$ -7.38\times 10^{-5} $&$ -5.04\times 10^{-4} $\\
$\lambda_5$&$ -1.58\times 10^{-3} $&$ 1.50\times 10^{-1} $\\
$\lambda_6$&$ -7.59\times 10^{-3} $&$ 3.06\times 10^{-2} $\\
$\lambda_7$&$ -4.23\times 10^{-1} $&$ 9.55\times 10^{-1} $\\
$\lambda_8$&$ 5.08 $&$ 1.16 $\\
$(Y_{27})_{11}$&$ -7.23\times 10^{-1} $&$ 1.87 $\\
$(Y_{27})_{12}$&$ 7.03\times 10^{-1} $&$ -1.09 $\\
$(Y_{27})_{22}$&$ -6.76\times 10^{-1} $&$ 6.30\times 10^{-1} $\\
$(Y_{\overline{351'}})_{11}$&$ -3.71\times 10^{-1} $&$ 7.39\times 10^{-1} $\\
$(Y_{\overline{351'}})_{12}$&$ 0 $&$ 0 $\\
$(Y_{\overline{351'}})_{22}$&$ 3.63\times 10^{-1} $&$ -2.58\times 10^{-1} $\\
\bottomrule
\end{longtable*}
\end{table}

\begin{table}[H]
\caption{Table of predictions for two example point in parameter space. All charge fermion masses are in units of $\mathrm{GeV}$, while the neutrino masses are in units $\mathrm{eV}$.}
\label{table:numeric-masses-and-angles}
\begin{longtable*}{rlrlrl}
\toprule
\multicolumn{1}{c}{quantity}&\multicolumn{1}{c}{experiment}&\multicolumn{2}{c}{parameter point 1}&\multicolumn{2}{c}{parameter point 2}\\
&\multicolumn{1}{c}{$y_i$}&\multicolumn{1}{c}{$f_i(x)$}&\multicolumn{1}{c}{$\chi_i$}&\multicolumn{1}{c}{$f_i(x)$}&\multicolumn{1}{c}{$\chi_i$}\\
\midrule
$m_c$&$0.236 $&$ 0.226 $&$ -0.432 $&$ 0.205 $&$ -1.30 $\\
$m_t$&$ 92.2 $&$ 94.0 $&$ +0.193 $&$ 105 $&$ +1.38 $\\
$\theta_{cb}$&$ 0.0409 $&$ 0.0358 $&$ -1.24 $&$ 0.0378 $&$ -0.757 $\\
$m_s$&$ 0.013 $&$ 0.0144 $&$ +1.06 $&$ 0.146 $&$ +1.20 $\\
$m_b$&$ 0.79 $&$ 0.79 $&$ +0.0021 $&$ 0.781 $&$ -0.112 $\\
$m_{\mu}$&$ 0.0599 $&$ 0.0613 $&$ +0.241 $&$ 0.0664 $&$ +1.08 $\\
$m_{\tau}$&$ 1.02 $&$ 0.867 $&$ -1.51 $&$ 1.03 $&$ +0.056$\\
$m_{\nu_3}$&&$0.135$&&$0.0824$&\\
$m_{\nu_2}$&&$0.126$&&$0.0659$&\\
$(m_{\nu_3}^2-m_{\nu_2}^2)10^{3}$&$ 2.32 $&$ 2.45 $&$ +0.568 $&$ 2.24 $&$ -0.325$\\
$\sin^2 \theta_{23}$&$ 0.386 $&$ 0.343 $&$ -1.12 $&$ 0.327 $&$ -1.52 $\\
\midrule
$\hat{\chi}^2$&&&\multicolumn{1}{l}{$\phantom{+}0.76$}&&\multicolumn{1}{l}{$\phantom{+}1.03$}\\
\bottomrule
\end{longtable*}
\end{table}

\section{Discussion\label{section:discussion}}

What we presented here is a model we believe is a good candidate for a minimal supersymmetric renormalizable
$E_6$ GUT. Let's see why this model can be considered as more minimal than our previous candidate \cite{Bajc:2013qra}.
Although the total number of degrees of freedom is now larger ($78>27+27$), what really counts are the number
of multiplets and, even more important, the total number of free parameters.
The number of parameters in the Higgs sector is now 11 complex minus 5 phases due to field redefinitions.
Yukawa sector adds another 3 real and 6 complex parameters. Together with one real gauge coupling we have
thus a total of $33$ real parameters. This is 7 more than in the minimal $\SO(10)$
\cite{Clark:1982ai,Aulakh:1982sw,Aulakh:2003kg}, but still $16$ less than
even in the simplified truncated version of $E_6$ in \cite{Bajc:2013qra}.

Is there any possible low-energy signature of this model? It is often said that $E_6$ could have possible
light extra generations, coming from remnants of the three copies of the $27_F$.
Light vector-like fermions of the SM group could indeed emerge out of $\EE$ if
they were associated with anomaly cancellation in a $\mathrm{TeV}$-scale extra $\mathrm{U}(1)$. In such a setup, the extra $\mathrm{U}(1)$ would be a linear combination
of the two $\mathrm{U}(1)$s present in the rank $6$ group $\EE$.  Phenomenology
of such $\mathrm{TeV}$ scale $\EE$ motivated models has been extensively studied,
see for eg.~\cite{rizzo}. In the present setup, however, there are no intermediate $\mathrm{U}(1)$s and any light vector-like states would be accidental.
In fact if we try to get such light states
from our solution, we find them hard to obtain. The question is, does the matrix for (as an example)
down quarks in eq.~\eqref{equation:Yukawa-terms-full} allow 4 or more zero eigenvalues once we
limit all $\bar v$'s to zero? This can in principle be obtained either by putting $c_2,f_4\to0$ (remember that
our solution already has $c_1=f_5=0$), or by imposing a vanishing determinant constraint to Yukawa matrices.
The first case points towards an SU(5) invariant vacuum\footnote{This can be easily seen from the expression for
$\lambda_2$ in (\ref{lambda2}): if $f_4\to0$, we need either $e_3\to0$ or $e_4\to0$; both lead to an SU(5) invariant
vacuum with $e_4,f_4\to0$, see \eqref{e4}.}, while the second one
constrains the Yukawa parameters and so a worse fit to data is expected. So we conclude that
such an extreme, albeit interesting situation is unlikely, at least in the given vacuum solution.
Although we cannot make the same conclusions in general, any possibility of light states, if available, will occur due to
fine-tuning in the superpotential parameters.

Another possibility for having light states could be to have flat directions. We checked by explicit computation that no such states are present in our solution. Although we omit the details of this computation here, the interested reader can reconstruct the mass matrix of the SM singlets via the all-singlet terms in the superpotential, which are given in equations~\eqref{equation:computed-invariant-first}--\eqref{equation:computed-invariant-last}. If the vacuum solution is then plugged-in, one discovers $4$ massless singlet states, all of which are in fact would-be Goldstone bosons. Note that the adjoint $78$ of $\EE$ contains $5$ SM singlets, while only $1$ remains unbroken among the SM generators, so the $4$ massless singlet states are the would-be Goldstone bosons eaten up by the $4$ broken singlet generators. We thus conclude there are no physical massless singlet states.
Other vacua could in principle be possible: one of them is described in Appendix~\ref{section:vacuum2}.

Proton decay is, as usual, quite hidden by details of superpartners' spectra: it is hard to disentangle the
GUT and SUSY breaking information from it. Once however colliders will (hopefully) tell us more about the
low energy spectrum, this $E_6$ theory as well as other grand unified theories could be tested better.
For the sake of completeness the forms of the low-energy $D=5$ operators are given in Appendix
\ref{Proton-decay}.

The model we presented here is the minimal known $E_6$, although for a more convincing proof we should satisfy
three more checks.

\begin{itemize}

\item
First, there is a possibility to redefine the charges under matter parity, so that the $27+\overline{27}$ parity
is now transferred to the fermionic sector. This means that the fields $351'+\overline{351'}+78$ should
alone break to the SM gauge group as well as allow for DT splitting. The fermionic sector would now consists of
$27^a_F$, $a=1,\ldots,4$ and $\overline{27}_F$. The Yukawa terms could be written schematically as

    \begin{align}
    \begin{pmatrix} 27^{a}_F\\ \overline{27}_F\end{pmatrix}^T
    \begin{pmatrix}
     Y_{\overline{351'}}^{ab}\langle \overline{351'}\rangle &m_{27}^a+\lambda^a\langle 78\rangle\\
    m_{27}^b+\lambda^b\langle 78\rangle& y_{351'}\langle 351'\rangle\\
    \end{pmatrix}
    \begin{pmatrix} 27^{b}_F\\ \overline{27}_F\end{pmatrix}.
    \end{align}

The total number of parameters is now the following: 5 complex parameters come from the Higgs superpotential,
3 phases of which can be rotated away by Higgs field redefinitions; 4 real diagonal components are given by the
only Yukawa matrix, and 8 complex and one real parameters are the off-diagonal terms; finally, 1 real gauge
coupling sums to a total of 29 real free parameters. This would be 4 real parameters less than the model in this
paper. Obviously there is no guarantee that such a model is realistic. We plan to come back to this issue in
the future.

\item
Second, we should study the complete three generation case, not only its two generation subsystem. Although
the number of parameters seems naively large enough, it is far from obvious that a successful fit is possible.
In fact already in the two generation case considered here we could not find a solution with vanishing $\chi^2$,
in spite of enough free parameters. However the  (at least partially) successful fit of the the minimal
SO(10) analogous case \cite{Babu:1992ia,Bajc:2001fe,Fukuyama:2002ch,Bajc:2002iw,Goh:2003sy,Goh:2003hf,
Bertolini:2004eq,Babu:2005ia,Bertolini:2006pe,Bajc:2008dc,Joshipura:2011nn,Altarelli:2013aqa,Dueck:2013gca}
with $10$ and $\overline{126}$ (instead of $27$ and $\overline{351'}$) Yukawa couplings make us feel optimistic.
Notice also that the usual obstruction of the neutrino mass either too low or unification violated
\cite{Aulakh:2005bd,Bajc:2005qe,Aulakh:2005mw,Bertolini:2006pe} is here avoided as shown in section
\ref{section:numeric-fit}.

\item
Third, the theory is not asymptotically free and has a huge gauge coupling beta function, more precisely $159$. This means,
similarly as in our previous $E_6$ model, or the minimal renormalizable supersymmetric $\SO(10)$, that a Landau pole
is close to the GUT scale and so the theory itself may be already in the non-perturbative regime. An indication
of problems being present already at the matching scale $M_{GUT}$ can be found in the calculation of the threshold corrections \cite{Aulakh:2011at,Aulakh:2013lxa}. The general problem of having large numbers of degrees of freedom (in our case through large representations) and the associated non-perturbativity is a problem, which is far from easy
to solve and well beyond the purpose of this paper, although some progress has been made recently
\cite{Abel:2008tx,Abel:2009bj} based on previous works on Seiberg dualities. We hope to come back to this
very interesting issue soon.

\end{itemize}

But even in the case the model presented here is the minimal one, other vacua could still be realistic
with in principle different predictions.

\section*{Acknowledgments}
The work of K.S.B is supported in part by the US Department of Energy Grant No. de-sc0010108.
The work of B.B. and V.S. is supported by the Slovenian Research Agency.

\appendix
\section{Gauge Boson masses \label{Gauge-Boson-masses}}
We compute in this appendix the expression for the gauge boson masses, so that we can confirm that solutions really break into the SM group. We write the mass terms in the Lagrangian as
\begin{align}
\mathcal{L}_{mass}&=g^2 A_\mu^{\;a} M^{ab} A^{\mu\,b},
\end{align}
\noindent
where $g$ is the $\EE$ coupling constant and the matrix $M^{ab}$ is computed via
\begin{align}
M^{ab}&:=\sum_i\mathrm{Tr}\left(\left(\hat{t}^{a} \langle\phi_i\rangle\right)^\dagger \left(\hat{t}^{b} \langle\phi_i\rangle\right)\right).
\end{align}
\noindent
The sum is over all representations containing VEVs (contributions come from the representations of the breaking sector), while $\hat{t}^a$ denotes the action of the $a$-th generator on the representation $\phi_i$. The mass matrix becomes block diagonal if we choose an appropriate basis (indices $a,b$), so that the basis gauge bosons have well defined transformation properties under the SM group (note that some of these states are complex). We omit the details of this calculation and only give the results, collected in Table~\ref{table:gauge-boson-masses}.

\def\THICK{0.2pt}
\begin{table}[H]
\caption{Masses-squared of gauge bosons in SM representations using the ansatz \hbox{$c_1=d_1=e_2=f_2=e_5=f_5=u_1=u_2=y=0$.}\label{table:gauge-boson-masses}}
\vskip 0.2cm
\centering
\scalebox{0.90}{
\begin{tabular}{p{1.5cm}p{1.5cm}p{2cm}p{8cm}}
\toprule
$\SO(10)\supset$&$\SU(5)\supset$&$\textrm{SM}\supset$&$\textrm{(mass)}^2/g^2$\\\midrule
$45$&$24$&$(8,1,\,0)$&$0$\\\addlinespace
$45$&$24$&$(1,3,\,0)$&$0$\\\addlinespace
$45$&$24$&$(1,1,\,0)$&$0$\\\midrule[\THICK]
$45$&$24$&$(3,2,+\tfrac{5}{6})$&$\frac{5}{6} \left(|\RED{e_4}|^2+|\RED{f_4}|^2\right)$\\
&&$(\overline{3},2,-\tfrac{5}{6})$&\\\midrule[\THICK]
$45$&$10$&$(3,2,+\tfrac{1}{6})$&$\frac{4}{15}|\RED{v}|^2+\frac{1}{2}|\RED{c_2}|^2+\frac{1}{2}|\RED{d_2}|^2$\\
&$\overline{10}$&$(\overline{3},2,-\tfrac{1}{6})$&$\quad+|\RED{e_1}|^2+|\RED{f_1}|^2+\frac{5}{6}|\RED{e_4}|^2+\frac{5}{6}|\RED{f_4}|^2$\\\midrule[\THICK]
$45$&$10$&$(\overline{3},1,-\tfrac{2}{3})$&$\frac{4}{15}|\RED{v}|^2+\frac{1}{2}|\RED{c_2}|^2+\frac{1}{2}|\RED{d_2}|^2+|\RED{e_1}|^2+|\RED{f_1}|^2$\\
&$\overline{10}$&$(3,1,+\tfrac{2}{3})$&\\\midrule[\THICK]
$45$&$10$&$(1,1,+1)$&$\frac{4}{15}|\RED{v}|^2+\frac{1}{2}|\RED{c_2}|^2+\frac{1}{2}|\RED{d_2}|^2+|\RED{e_1}|^2+|\RED{f_1}|^2$\\
&$\overline{10}$&$(1,1,-1)$&\\\midrule[\THICK]
$16$&$10$&$(3,2,+\tfrac{1}{6})$&$\frac{1}{60} |\sqrt{15} \RED{w}-\RED{v}|^2+|\RED{e_3}|^2+|\RED{f_3}|^2+\frac{1}{2}|\RED{e_4}|^2+\frac{1}{2}|\RED{f_4}|^2$\\
$\overline{16}$&$\overline{10}$&$(\overline{3},2,-\tfrac{1}{6})$&\\\midrule[\THICK]
$16$&$10$&$(\overline{3},1,-\tfrac{2}{3})$&$\frac{1}{60} |\sqrt{15} \RED{w}-\RED{v}|^2+|\RED{e_3}|^2+|\RED{f_3}|^2+\frac{1}{2}|\RED{e_4}|^2+\frac{1}{2}|\RED{f_4}|^2$\\
$\overline{16}$&$\overline{10}$&$(3,1,+\tfrac{2}{3})$&\\\midrule[\THICK]
$16$&$10$&$(1,1,+1)$&$\frac{1}{60} |\sqrt{15} \RED{w}-\RED{v}|^2+|\RED{e_3}|^2+|\RED{f_3}|^2+\frac{1}{2}|\RED{e_4}|^2+\frac{1}{2}|\RED{f_4}|^2$\\
$\overline{16}$&$\overline{10}$&$(1,1,-1)$&\\\midrule[\THICK]
$16$&$\phantom{0}\overline{5}$&$(\overline{3},1,+\tfrac{1}{3})$&$\frac{1}{4} |\RED{w}+\sqrt{3/5}\,\RED{v}|^2+\frac{1}{2}|\RED{c_2}|^2+\frac{1}{2}|\RED{d_2}|^2+|\RED{e_1}|^2+|\RED{f_1}|^2+$\\
$\overline{16}$&$\phantom{0}5$&$(3,1,-\tfrac{1}{3})$&$\qquad\qquad +|\RED{e_3}|^2+|\RED{f_3}|^2+\frac{1}{2}|\RED{e_4}|^2+\frac{1}{2}|\RED{f_4}|^2$\\\midrule[\THICK]
$16$&$\phantom{0}\overline{5}$&$(1,2,-\tfrac{1}{2})$&$\frac{1}{4} |\RED{w}+\sqrt{3/5}\,\RED{v}|^2+\frac{1}{2}|\RED{c_2}|^2+\frac{1}{2}|\RED{d_2}|^2+|\RED{e_1}|^2+|\RED{f_1}|^2+$\\
$\overline{16}$&$\phantom{0}5$&$(1,2,+\tfrac{1}{2})$&$\qquad\qquad +|\RED{e_3}|^2+|\RED{f_3}|^2+\tfrac{1}{2}|\RED{e_4}|^2+\tfrac{1}{2}|\RED{f_4}|^2$\\\midrule[\THICK]
$45$\par $\phantom{0}1$&$\phantom{0}1$\par $\phantom{0}1$&$(1,1,\,0)$\par $(1,1,\,0)$&
They mix:\vskip -0.6cm
\begin{equation*}
\tfrac{2}{3} \Big((A+B)\pm\sqrt{(A+B)^2-\tfrac{15}{4}A B}\Big),
\end{equation*}
\vskip -0.3cm
$A\equiv 4 |\RED{e_3}|^2 + 4 |\RED{f_3}|^2 +|\RED{e_4}|^2+ |\RED{f_4}|^2$\par $B\equiv 4 |\RED{e_1}|^2+4|\RED{f_1}|^2+|\RED{c_2}|^2+|\RED{d_2}|^2$
\\\midrule[\THICK]
$16$\par $\overline{16}$&$\phantom{0}1$\par $\phantom{0}1$&$(1,1,\,0)$\par $(1,1,\,0)$&
They mix:\vskip -0.6cm
\begin{equation*}
\tfrac{1}{2}\Big((C+D+|F|^2)\pm \sqrt{(C-D)^2+16|E|^2}\;\Big),
\end{equation*}
\vskip -0.3cm
$C\equiv |\RED{c_2}|^2+2 |\RED{e_1}|^2+2 |\RED{f_3}|^2+|\RED{e_4}|^2$\par
$D\equiv |\RED{d_2}|^2+2 |\RED{f_1}|^2+2 |\RED{e_3}|^2+ |\RED{f_4}|^2$\par
$E\equiv \RED{e_1} \RED{e_3}^\ast + \RED{f_1}^\ast \RED{f_3}$\par
$F\equiv \sqrt{\tfrac{5}{6}} \RED{v}-\sqrt{\tfrac{1}{2}}\RED{w}$
\\\bottomrule
\end{tabular}
}
\end{table}

\section{Particle identification \label{Particle-identification}}

{\small
\begin{table}[H]
\caption{Identification of doublets and triplets in the representations of the Higgs sector.\label{table:labels-doublets-triplets}}
\begin{longtable*}{llll@{\hspace{0.5cm}}p{4cm}}
\toprule
label&{\footnotesize $E_6\supseteq\SO(10)\supseteq\SU(5)$}&label&{\footnotesize $E_6\supseteq\SO(10)\supseteq\SU(5)$}&doublet\par triplet\\\midrule
$\overline{D}_0,\overline{T}_0$&$\phantom{0}78\phantom{'}\supseteq\phantom{0}16\supseteq\phantom{0}\overline{5}$&$D_0,T_0$&$\phantom{0}78\phantom{'}\supseteq\phantom{0}\overline{16}\supseteq\phantom{0} 5$&$\tfrac{1}{\sqrt{12}}(t_L^6\pm i\, t_L^7)$\par $\tfrac{1}{\sqrt{12}}\bar{t}_\alpha{}^{31},\tfrac{1}{\sqrt{12}}t^\alpha{}_{31}$\\\addlinespace
$D_1,T_1$&$\phantom{0}27\phantom{'}\supseteq\phantom{0}10\supseteq\phantom{0}5$&$\overline{D}_1,\overline{T}_1$&$\phantom{0}\overline{27}\phantom{'}\supseteq\phantom{0}10\supseteq\phantom{0}\overline{5}$&$L'^c$\par $d'$\\\addlinespace
$\overline{D}_2,\overline{T}_2$&$\phantom{0}27\phantom{'}\supseteq\phantom{0}10\supseteq\phantom{0}\overline{5}$&$D_2,T_2$&$\phantom{0}\overline{27}\phantom{'}\supseteq\phantom{0}10\supseteq\phantom{0}5$&$L'$\par $d'^c$\\\addlinespace
$\overline{D}_3,\overline{T}_3$&$\phantom{0}27\phantom{'}\supseteq\phantom{0}16\supseteq\phantom{0}\overline{5}$&$D_3,T_3$&$\phantom{0}\overline{27}\phantom{'}\supseteq\phantom{0}\overline{16}\supseteq\phantom{0}5$&$L$\par $d^c$\\\addlinespace
$D_4,T_4$&$351'\supseteq\phantom{0}10\supseteq\phantom{0}5$&$\overline{D}_4,\overline{T}_4$&$\overline{351'}\supseteq\phantom{0}10\supseteq\phantom{0}\overline{5}$&$Q d^c-L e^c-4L'^c \nu^c$\par $QL-u^c d^c-4d's$\\\addlinespace
$\overline{D}_5,\overline{T}_5$&$351'\supseteq\phantom{0}10\supseteq\phantom{0}\overline{5}$&$D_5,T_5$&$\overline{351'}\supseteq\phantom{0}10\supseteq\phantom{0} 5$&$Q u^c-L\nu^c-4L' s$\par $u^c e^c-d^c\nu^c+QQ-4d'^c s$\\\addlinespace
$\overline{D}_6,\overline{T}_6$&$351'\supseteq\phantom{0}16\supseteq\phantom{0}\overline{5}$&$D_6,T_6$&$\overline{351'}\supseteq\phantom{0}\overline{16}\supseteq\phantom{0}5$&$-L s$\par $-d^c s$\\\addlinespace
$\overline{D}_7,\overline{T}_7$&$351'\supseteq 126\supseteq\phantom{0}\overline{5}$&$D_7,T_7$&$\overline{351'}\supseteq \overline{126}\supseteq\phantom{0} 5$&$-Qu^c-3L\nu^c$\par $-u^c e^c-3 d^c\nu^c-QQ$\\\addlinespace
$D_8,T_8$&$351'\supseteq126\supseteq 45$&$\overline{D}_8,\overline{T}_8$&$\overline{351'}\supseteq\overline{126}\supseteq \overline{45}$&$Qd^c+3Le^c$\par $QL+u^c d^c$\\\addlinespace
$D_9,T_9$&$351'\supseteq\overline{144}\supseteq\phantom{0}5$&$\overline{D}_9,\overline{T}_9$&$\overline{351'}\supseteq 144\supseteq\phantom{0}\overline{5}$&$-Qd'^c+4L'^c\nu^c+L'e^c$\par $-QL'+u^c d'^c+4d' \nu^c$\\\addlinespace
$\overline{D}_{10},\overline{T}_{10}$&$351'\supseteq\overline{144}\supseteq\phantom{0}\overline{5}$&$D_{10},T_{10}$&$\overline{351'}\supseteq 144\supseteq\phantom{0} 5$&$-L'\nu^c$\par $-d'^c\nu^c$\\\addlinespace
$D_{11},T_{11}$&$351'\supseteq\overline{144}\supseteq 45$&$\overline{D}_{11},\overline{T}_{11}$&$\overline{351'}\supseteq 144\supseteq\overline{45}$&$-dd'^c-3e'e^c$\par $-QL'-u^c d'^c$\\\addlinespace
$\phantom{D_{1}}\overline{T}_{12}$&$351'\supseteq 126\supseteq \overline{50}$&$\phantom{D_{1}}T_{12}$&$\overline{351'}\supseteq \overline{126}\supseteq 50$&$/$\par $2u^c e^c- QQ$\\\bottomrule
\end{longtable*}
\end{table}
}

In this appendix, we write the definitions of various states needed in the paper. The SM singlet VEVs were already defined in Table~\ref{table:singlet-labels}. We supplement the list of definitions with Tables~\ref{table:labels-doublets-triplets} and \ref{table:weak-triplets}. In Table~\ref{table:labels-doublets-triplets} we define the doublets $(1,2,+1/2)$, antidoublets $(1,2,-1/2)$, triplets $(3,1,-1/3)$ and antitriplets $(\bar{3},1,+1/3)$; these definitions are needed for DT splitting. In Table~\ref{table:weak-triplets}, we define the weak triplets $(1,3,\pm 1)$ relevant for type II seesaw. All the states in the two-index $351'$ can be specified by writing the basis states of this representation by using two labels of the fundamental $27$. In this notation, both labels are SM representation in the $27$, assumed to have all the color and weak indices contracted in the correct manner to obtain the desired SM state in the $351'$. More details on this notation can be found in~\cite{Bajc:2013qra}.

\begin{table}[H]
\caption[Seesaw type II triplet scalars]{Induced VEVs in weak triplet scalars $(1,3,\pm 1)$ leading to seesaw type II.\label{table:weak-triplets}}
\vskip 0.2cm
\centering
\begin{tabular}{llp{1cm}@{\hspace{2cm}}llp{1cm}}
\toprule
label&{\footnotesize $E_6\supseteq\SO(10)\supseteq\SU(5)$}&p.n.&label&{\footnotesize $E_6\supseteq\SO(10)\supseteq\SU(5)$}&state\\\midrule
$\overline{\Delta}_1$&$351'\supseteq 126\supseteq\phantom{0}\overline{15}$&$L\phantom{'}L$&$\Delta_1$&$\overline{351'}\supseteq\overline{126}\supseteq 15$&$\bar{L}\phantom{'}\bar{L}$\\
$\overline{\Delta}_2$&$351'\supseteq\overline{144}\supseteq\phantom{0}\overline{15}$&$L\phantom{'}L'$&$\Delta_2$&$\overline{351'}\supseteq 144\supseteq 15$&$\bar{L}\phantom{'}\bar{L}'$\\
$\overline{\Delta}_3$&$351'\supseteq\phantom{0}54\supseteq\phantom{0}\overline{15}$&$L'L'$&$\Delta_3$&$\overline{351'}\supseteq\phantom{0}54\supseteq 15$&$\bar{L}'\bar{L}'$\\
$\Delta_4$&$351'\supseteq\phantom{0}54\supseteq\phantom{0}15$&$L'^c L'^c$&$\overline{\Delta}_4$&$\overline{351'}\supseteq\phantom{0}54\supseteq\overline{15}$&$\bar{L}'^c \bar{L}'^c$\\\bottomrule
\end{tabular}
\end{table}

\section{DT splitting and Goldstone modes\label{section:simplified-DT}}

Analysis of the DT splitting in the group $\EE$ is complicated by the fact that a vacuum breaking to the SM group will automatically cause a doublet and a triplet mode to be massless. These massless modes are would-be Goldstone bosons: $\EE\to\mathrm{SM}$ causes the breaking of $78-12=66$ generators, with a doublet-antidoublet and triplet-antitriplet pair among them. One possible procedure to compute the condition for an extra massless mode in a matrix $M$, for which $\mathrm{det}\,M=0$, is to take
\begin{align}
\frac{\lim_{\varepsilon\to 0}\big(\det(M-\varepsilon I)/\varepsilon\big)}{\langle e|f\rangle}&=0,
\end{align}
where $e$ and $f$ are the already present left and right null-eigenvectors of $M$, respectively. We present below, however, a simplified procedure of computing the conditions of DT splitting in the presence of a Goldstone mode. Its advantages are that it is computationally less intensive and that no issues with singularities, such as $\langle e|f\rangle$=0, arise in the procedure.

Suppose we use a generic label $A$ for an $n\times n$ complex matrix. Although the true scalar mass-squared matrix is in fact the hermitian and positive definite matrix $A^\dagger A$, it is more efficient to work with $A$. $A$ might not necessarily be diagonalizable,
but it has a singular value decomposition. The presence of a zero eigenmode in $A^\dagger A$ implies
\begin{align}
\mathrm{det}\,A&=0.
\end{align}
We will rotate this matrix into a basis, where the left and right Goldstone modes correspond to the first basis vector of the rows and columns, respectively. We first write $A$ in  \hbox{$(1+(n-1))\times (1+(n-1))$} block form:
\begin{align}
A&=\begin{pmatrix} m&\mathbf{m_R}^\dagger\\ \mathbf{m_L}&\mathbf{M}\\\end{pmatrix},
\end{align}
where boldface small letters denote $n-1$ column vectors, and boldface capital letters denote matrices. Since $A$ has a massless mode, there exist left and right null-eigenvectors $\vec{e}_0$ and $\vec{f}_0$, respectively:
\begin{align}
A^\dagger\vec{e}_0&=0,&A\vec{f}_0&=0.\label{null-eq}
\end{align}
Providing we choose the phases such that the first components of $\vec{e}_0$ and $\vec{f}_0$ are positive, we
define $(n-1)$ column vectors $\mathbf{e}$ and $\mathbf{f}$ via
\begin{align}
\frac{\vec{e}_0}{|\vec{e}_0|}&=:\begin{pmatrix} \sqrt{1-\mathbf{e}^\dagger\mathbf{e}}\\\mathbf{e}\\\end{pmatrix}&
\frac{\vec{f}_0}{|\vec{f}_0|}&=:\begin{pmatrix} \sqrt{1-\mathbf{f}^\dagger\mathbf{f}}\\\mathbf{f}\\\end{pmatrix}.
\end{align}
Writing the null-eigenvector conditions in equations~\eqref{null-eq} in block form, we get $3$ independent equations (two $(n-1)$-vector, $1$ scalar):
\begin{align}
\sqrt{1-\mathbf{e}^\dagger\mathbf{e}}\;\mathbf{m_R}+\mathbf{M}^\dagger\,\mathbf{e}&=0,\\
\sqrt{1-\mathbf{f}^\dagger\mathbf{f}}\;\mathbf{m_L}+\mathbf{M}^{\phantom{\dagger}}\,\mathbf{f}&=0,\\
m\sqrt{1-\mathbf{e}^\dagger\mathbf{e}}\sqrt{1-\mathbf{f}^\dagger\mathbf{f}}-\mathbf{e}^\dagger\,\mathbf{M}\,\mathbf{f}&=0
\end{align}
These can for example be used to define $m$, $\mathbf{m_L}$ and $\mathbf{m_R}$ in terms of $\mathbf{M}$, $\mathbf{e}$ and $\mathbf{f}$. We now define a $n\times n$ unitary matrix $U(\mathbf{x})$, whose form will be useful for rotating the basis of $A$: in block form, $U$ is written as
\begin{align}
U(\mathbf{x}):=\begin{pmatrix} \sqrt{1-\mathbf{x}^\dagger\mathbf{x}}& \mathbf{x}^\dagger\\ -\mathbf{x}&\Lambda(\mathbf{x})\\\end{pmatrix},
\end{align}
\noindent
where
\begin{align}
\Lambda(\mathbf{x}):=I-\frac{\mathbf{x}\,\mathbf{x}^\dagger}{1+\sqrt{1-\mathbf{x}^\dagger \mathbf{x}}}.
\end{align}
It is possible to check explicitly that $U$ is indeed unitary and that the following relations hold:
\begin{align}
U(\mathbf{x})\,U(\mathbf{x})^\dagger&=I,\\
U(\mathbf{x})^{-1}=U(\mathbf{x})^\dagger&=U(-\mathbf{x}),\\
\Lambda(\mathbf{x})^\dagger&=\Lambda(\mathbf{x})=\Lambda(-\mathbf{x}),\\
\Lambda(\mathbf{x})^{-1}&=I+\frac{\mathbf{x}\,\mathbf{x}^\dagger}{\sqrt{1-\mathbf{x}^\dagger\mathbf{x}}\;(1+\sqrt{1-\mathbf{x}^\dagger\mathbf{x}})},\\
\mathrm{det}\Lambda(\mathbf{x})&=1-\frac{\mathbf{x}^\dagger \mathbf{x}}{1+\sqrt{1-\mathbf{x}^\dagger\mathbf{x}}}.\label{equation:lambda-det}
\end{align}
The properly rotated matrix $A$, denoted by $A'$, is then by explicit computation equal to
\begin{align}
A'&:=U(\mathbf{e})\,A\,U(\mathbf{f})^{-1}=\begin{pmatrix} 0&\mathbf{0}^\dagger\\ \mathbf{0}& \mathbf{M_{rot}}\\\end{pmatrix},
\end{align}
\noindent
where
\begin{align}
\mathbf{M_{rot}}&=\Lambda(\mathbf{e})^{-1}\mathbf{M}\,\Lambda(\mathbf{f})^{-1}.
\end{align}
We have indeed rotated into a basis, where the first column and row correspond to the zero eigenmodes. Crucially, the formula for the rotated \hbox{$(n-1)\times(n-1)$} block has only one term, where a simple biunitary rotation is performed on the original block $\mathbf{M}$; another zero-eigenmode can now be simply imposed by taking $\mathrm{det}\mathbf{M_{rot}}=0$. But since we now have
\begin{align}
\mathrm{det}\mathbf{M}&=\mathrm{det}\mathbf{M_{rot}}\;\mathrm{det}\big(\Lambda(\mathbf{e})\Lambda(\mathbf{f})\big).
\end{align}
 it is sufficient to impose $\det\mathbf{M}=0$. The only possible caveat is the possibility that either $\mathrm{det}\,\Lambda(\mathbf{e})=0$ or $\mathrm{det}\,\Lambda(\mathbf{f})=0$; considering equation~\eqref{equation:lambda-det} and that $\mathbf{e}$ and $\mathbf{f}$ are parts of normalized vectors, this can happen only if $\mathbf{e}^\dagger\mathbf{e}=1$ or $\mathbf{f}^\dagger\mathbf{f}=1$, which would imply that the zero modes of $A$ have a zero component in the direction of the first vector of the original basis. But since the eigenvalues of $A^\dagger A$ (the determinant of $A^\dagger A-\lambda I$) do not change if we rearrange the rows or the columns of $A$, we can always rearrange the original basis so that we take the $i$-th row and the $j$-th column to be the preferred one for the left and right null-eigenmodes, respectively.

The main result thus states the following: if an $n\times n$ complex matrix $A$ has a zero mode present, an additional zero mode is obtained by demanding the $(i,j)$-th minor of $A$ to vanish (the subdeterminant of $A$, when the $i$-th row and $j$-th column are removed), where the indices $i$ and $j$ can be arbitrarily chosen, as long as the left and right null-eigenvectors of $A$ have respectively a nonzero $i$-th and $j$-th component.

\section{An alternative vacuum\label{section:vacuum2}}
The Yukawa fit presented in this paper is based on the vacuum computed in section~\ref{section:SSB}. Since we have not been able to obtain (yet) a full classification of all vacua in this $\EE$ model, we cannot conclude much about the broader possibilities of suitable vacua (especially the Yukawa sector), but we were able to find one other alternative Standard Model vacuum. We can obtain it by taking the following ansatz of vanishing VEVs:
\begin{align}
c_2=d_2=e_5=f_5=e_2=f_2=u_1=u_2=y&=0.\label{equation:alt-ansatz}
\end{align}
This ansatz is similar to the ansatz of the original vacuum, but the vanishing of $e_5,f_5$ is now paired up with the vanishing of $c_2,d_2$ instead of $c_1,d_1$. For the purposes of this appendix, we omit the specific form of the solutions, as well as other details, such as the check that the unbroken group is indeed that of the Standard Model.

Looking at the Yukawa terms in equation~\eqref{equation:Yukawa-terms-full}, we see that the alternative vacuum with $c_2=f_5=0$ decouples the $16_F$ from the $10_F$ of $\SO(10)$ (in leading order of $m_{EW}/M_{GUT}$). The heavy vector-like exotics (in the down-quark sector and charged lepton sector) in the fermionic $27_F$ are thus purely in the $10_F$ part. The low-energy mass matrices are thus simply those for the $16_F$, the analysis of the Yukawa sector becomes linear, and numerically one can make use of the fit for the minimal supersymmetric $\SO(10)$ model~\cite{Babu:1992ia,Bajc:2001fe,Fukuyama:2002ch,Bajc:2002iw,Goh:2003sy,Goh:2003hf,Bertolini:2004eq,Babu:2005ia,Bertolini:2006pe,Bajc:2008dc,Joshipura:2011nn,Altarelli:2013aqa,Dueck:2013gca} with the Higgs in the $10$ and $\overline{126}$ coupling to fermion pairs in $16_F^2$.

The intriguing possibility of this alternative vacuum, which recovers the $\SO(10)$ limit, is somewhat marred by issues in DT splitting. Given the ansatz in equation~\eqref{equation:alt-ansatz}, the doublet and triplet mass matrices become block diagonal with the following block form (with the basis of the barred states rearranged in the same order as for unbarred):
\begin{align}
\begin{bmatrix}D_0 &D_3 &D_6 &D_9 &D_{10} &D_{11}\\\end{bmatrix},\quad \begin{bmatrix}D_1 &D_2 &D_4 &D_5\\\end{bmatrix},\quad \begin{bmatrix}D_7 &D_8\\\end{bmatrix},\\
\begin{bmatrix}T_0 &T_3 &T_6 &T_9 &T_{10} &T_{11}\\\end{bmatrix},\quad \begin{bmatrix}T_1 &T_2 &T_4 &T_5\\\end{bmatrix},\quad \begin{bmatrix}T_7 &T_8& T_{12}\\\end{bmatrix}.
\end{align}
The Goldstone modes, for example, turn out to be in the first block. The problem now is, however, that a fine-tuning is block specific: the light Higgs lives only in one of the blocks. We see from equation~\eqref{equation:Yukawa-terms-full} that the low-energy fermionic mass matrices $M_D^T$ and $M_E$ are now controlled solely by the EW VEVs $\bar{v}_2$, $\bar{v}_4$ and $\bar{v}_8$. In the $\bar{v}_8=0$ case, we get the unwanted mass relation $M_D^T=M_E$, but $\overline{D}_{8}$ is in a separate block compared to $\overline{D}_2$ and $\overline{D}_4$. A realistic pattern of fermion masses would thus require a double fine-tuning: one in the second block and one in the third block. Due to this feature, we consider this vacuum to be of less interest: beside the aesthetically unpleasing extra fine-tuning, the additional light Higgs pair $H'_u+H'_d$ pair gives large threshold corrections in the running of the gauge couplings, possibly spoiling unification. The model contains many heavy states though, so the situation regarding the RGE is not clear-cut. For determining the feasibility of this vacuum, further investigation would be necessary.

We conclude this section of the Appendix with a brief elaboration on which kind of ansatz is suitable for a good vacuum, i.e.~we motivate equations~\eqref{equation:ansatz-first}--\eqref{equation:ansatz-last} and \eqref{equation:alt-ansatz}. The considerations will be very similar to the ones in the $\EE$ breaking sector, where the $78$ is omitted~\cite{Bajc:2013qra} and a full classification of vacua is known.
Suppose we look for a specific vacuum solution: we want it to be as simple as possible (it has as many vanishing VEVs as possible), yet it needs to be able to break $\EE$ to the SM group. Due to the $D$-terms in SUSY, we assume a conjugate-symmetric ansatz, where a vanishing VEV in $27$ or $351'$ implies that the corresponding (conjugate) VEV in the $\overline{27}$ or $\overline{351'}$ also vanishes, and vice versa. First, we identify the $\SU(5)$ breaking VEVs from Table~\ref{table:singlet-labels}: $e_4$, $e_5$, $f_4$, $f_5$ and $y$. The $F_y$ equation of motion automatically implies $y=0$. Since $\SU(5)$ needs to be broken, either the pair $e_4$, $f_4$ is non-vanishing, or the pair $e_5$, $f_5$. Considerations of alignment symmetry in~\cite{Bajc:2013qra} imply that the choice is irrelevant, since these pairs are exchanged if one changes the embedding of $\SO(10)$ in $\mathrm{E}_6$, such that the two $\bar{5}$s of $\SU(5)$ in the $27$ are exchanged. Therefore one pair needs to be necessarily non-zero, while we can try a vanishing ansatz for the other pair. In the $27+\overline{27}$ part of the breaking sector, the pairs $c_1$, $d_1$ and $c_2$, $d_2$ also get exchanged under alignment symmetry; we assume one pair to be non-zero (such that the $27+\overline{27}$ part does indeed contribute to the symmetry breaking), but we can again try setting the other pair to vanish in the simple ansatz. The choice of the vanishing pair now points to either the scenario of the main vacuum of this paper, or the the alternative vacuum presented in this appendix. The remaining part of the ansatz, $e_2=f_2=0$ (which transforms into itself under alignment symmetry), is suggested from the solutions in the model without the $78$, while $u_1=u_2=0$ then follows as a consequence of the $F$-terms. We finish the ansatz discussion with the following points:
\begin{itemize}
\item The main vacuum solution of the paper follows from the ansatz, which is a direct extension of the solution ansatz in the absence of the $78$~\cite{Bajc:2013qra}. There, the EOMs were simpler, and this ansatz actually represented the most general SM solution once gauge freedom and the $F$-term equations were accounted for. Note that only the ansatz can be extended, the main solution itself is not merely an extension of the solution when the $78$ is omitted.
\item The alternative ansatz leads to a SM vacuum only after the $78$ was included; there is no such option if the $78$ is omitted.
\item A complete classification would tell us, whether still other nonequivalent vacua exist, where the VEVs are non-vanishing. If the $78$ is omitted, such vacua did not exist, but we expect this situation to change due to more terms and more VEVs in the EOM.
\end{itemize}

\section{Proton decay \label{Proton-decay}}
For completeness let's summarize the analysis of $D=5$ proton decay in this model, similar to the analysis done in~\cite{Bajc:2013qra}, obtaining analogous results. The low-energy operators in the superpotential, which are relevant for proton decay, are
\begin{align}
W\big|_{\textrm{proton}}&=-\Big[\big(\overline{C}_1^{inA}-\overline{C}_1'^{imA}(X^T)_{m}{}^{n}\big)\big[(1+X^\ast X^T)^{-1/2}\big]_{n}{}^{j}\;(\hat{\mathcal{M}}_T^{-1})_{AB}\; C_1^{klB}\Big]\;Q_i \hat{L}_j Q_k Q_l\nonumber\\
&\quad -\Big[\big(\overline{C}_2^{njA}+\tfrac{2}{3}\overline{C}_2'^{mjA}(X^T)_{m}{}^{n}\big)\big[(1+\tfrac{4}{9}X^\ast X^T)^{-1/2}\big]_{n}{}^{i}\;(\hat{\mathcal{M}}_T^{-1})_{AB}\; C_2^{klB}\Big]\;\hat{d}^c_i u^c_j u^c_k e^c_l,
\end{align}
\noindent
with the $X$ defined in equation~\eqref{equation:X0-definition} and the mass matrix of $\mathcal{M}_{\textrm{T}}$ as already defined in equation~\eqref{equation:DT-matrix} of section~\ref{section:DT-splitting}. Note that the $\mathcal{M}_{\textrm{T}}$ has a zero eigenmode corresponding to the would-be Goldstone, so its inverse cannot be directly computed; we instead write $\hat{\mathcal{M}}_{T}^{-1}=\lim_{M\to\infty}(\mathcal{M}_T+M\;e\,f^T)^{-1}$, with $e$ and $f$ being the left and right column-eigenvectors of $\mathcal{M}_T$, respectively.

The $C$ coefficients are computed to be

\begin{align}
2\;C_1^{ijA}&=-\yd^{ij}\,\delta^{A}{}_{1}+\tfrac{1}{2\sqrt{10}}\,\yt^{ij}\,\delta^{A}{}_{5}
-\tfrac{1}{2\sqrt{6}}\,\yt^{ij}\,\delta^{A}{}_{7}-\tfrac{1}{2\sqrt{3}}\,\yt^{ij}\,\delta^{A}{}_{12},\\
2\;C_2^{ijA}&=-\yd^{ij}\,\delta^{A}{}_{1}+\tfrac{1}{2\sqrt{10}}\,\yt^{ij}\,\delta^{A}{}_{5}
-\tfrac{1}{2\sqrt{6}}\,\yt^{ij}\,\delta^{A}{}_{7}+\tfrac{2}{2\sqrt{3}}\,\yt^{ij}\,\delta^{A}{}_{12},\\
2\;\overline{C}_1^{ijA}&=-\yd^{ij}\,\delta^{A}{}_{2}+
\tfrac{1}{2\sqrt{10}}\,\yt^{ij}\,\delta^{A}{}_{4}+\tfrac{1}{2\sqrt{2}}\,\yt^{ij}\,\delta^{A}{}_{8},\\
2\,\overline{C}_1'^{ijA}&=\phantom{-}\yd^{ij}\,\delta^{A}{}_{3}-
\tfrac{1}{2\sqrt{10}}\,\yt^{ij}\,\delta^{A}{}_{9}-\tfrac{1}{2\sqrt{2}}\,\yt^{ij}\,\delta^{A}{}_{11},\\
2\;\overline{C}_2^{ijA}&=-\yd^{ij}\,\delta^{A}{}_{2}+
\tfrac{1}{2\sqrt{10}}\,\yt^{ij}\,\delta^{A}{}_{4}-\tfrac{1}{2\sqrt{2}}\,\yt^{ij}\,\delta^{A}{}_{8},\\
2\,\overline{C}_2'^{ijA}&=\phantom{-}\yd^{ij}\,\delta^{A}{}_{3}-
\tfrac{1}{2\sqrt{10}}\,\yt^{ij}\,\delta^{A}{}_{9}+\tfrac{1}{2\sqrt{2}}\,\yt^{ij}\,\delta^{A}{}_{11}.
\end{align}

We see that the $C$-coefficients are the same as the coefficients in~\cite{Bajc:2013qra}, if we cross out the contributions from the extra Yukawa term in that model. More specifically, notice that there are no $A=0$ contributions from triplets/antitriplets in the new representation $78$, since the $78$ is not present in the Yukawa sector (it does not couple to two $27_F$'s).

Once the model parameters are fit to the experimental values of the fermion masses and mixings, the four-fermion amplitude
mediating proton decay is fixed and thus potentially dangerous, but it can always be suppressed by implementing
a split supersymmetric scenario without changing any other conclusion.

\end{document}